\documentclass[twocolumn]{aastex631}
\usepackage[dvipsnames]{xcolor}
\usepackage{amsmath}
\usepackage{booktabs}
\usepackage{float}
\usepackage{ragged2e}
\usepackage{CJK}
\usepackage{soul}

\newcommand{\astrid}{\textsc{ASTRID}}
\newcommand{\brahma}{\textsc{BRAHMA}}
\newcommand{\Abrahma}{\textsc{AMBRA}}
\newcommand{\Msun}{$M_{\odot}$}
\newcommand{\gnz}{\textsc{GN-z11}}
\newcommand{\ceers}{\textsc{CEERS-1019}}

\shorttitle{Early MBH Growth in AMBRA}
\shortauthors{Y.Zhou et al.}

\begin{document}
\begin{CJK*}{UTF8}{bsmi}

%\title{First results of ASTRID-BRAHMA \tiziana{we need to change name- too confusing otherwise}: Seed mergers boost high-z massive black hole growth}

%\title{A Pathway to the First Massive Black Holes: Abundant Seeds and Early Mergers in Rare Environments}
%\title{First results of \Abrahma: A Pathway to the First Massive Black Holes}
\title{First results of \Abrahma: Abundant Seeds and Early Mergers as a Pathway to the First Massive Black Holes}

\author[0000-0002-8828-8461]{Yihao Zhou (周亦豪)}
\affiliation{McWilliams Center for Cosmology, Department of Physics, Carnegie Mellon University, Pittsburgh, PA 15213, USA}

\author[0000-0002-7080-2864]{Aklant Kumar Bhowmick}
\affiliation{University of Virginia 530 McCormick Rd Charlottesville, VA 22904, USA}
\affiliation{Virginia Institute for Theoretical Astronomy, University of Virginia, Charlottesville, VA 22904, USA}
\affiliation{The NSF-Simons AI Institute for Cosmic Origins, USA}
\author[0000-0002-6462-5734]{Tiziana Di Matteo}
\affiliation{McWilliams Center for Cosmology, Department of Physics, Carnegie Mellon University, Pittsburgh, PA 15213, USA}
\author[0009-0006-7511-0329]{Patrick LaChance}
\affiliation{McWilliams Center for Cosmology, Department of Physics, Carnegie Mellon University, Pittsburgh, PA 15213, USA}
\author[0000-0003-0697-2583]{Rupert Croft}
\affiliation{McWilliams Center for Cosmology, Department of Physics, Carnegie Mellon University, Pittsburgh, PA 15213, USA}
\author[0000-0002-2183-1087]{Laura Blecha}
\affiliation{Department of Physics, University of Florida, Gainesville, FL 32611, USA}
\author[0000-0001-5803-5490]{Simeon Bird}
\affiliation{Department of Physics \& Astronomy, University of California, Riverside, 900 University Ave., Riverside, CA 92521, USA}
\author[0000-0002-5653-0786]{Paul Torrey}
\affiliation{University of Virginia 530 McCormick Rd Charlottesville, VA 22904, USA}
\affiliation{Virginia Institute for Theoretical Astronomy, University of Virginia, Charlottesville, VA 22904, USA}
\affiliation{The NSF-Simons AI Institute for Cosmic Origins, USA}
\author{Lars Hernquist}
\affiliation{Harvard-Smithsonian Center for Astrophysics, 60 Garden Street, Cambridge, MA 02138, USA}

\correspondingauthor{Yihao Zhou}
\email{yihaoz@andrew.cmu.edu}


\begin{abstract}

%JWST observations are revealing massive black holes in galaxies at $z\gtrsim9$, posing a challenge for models of early black hole assembly.
We present the first results from \Abrahma\ simulation (\textbf{A}STRID with \textbf{M}BH seeding from \textbf{BRA}HMA) evolved to $z=8$.
%with a model for black hole seeding based on the gas properties of host galaxies.
%\Abrahma\ implements physically motivated black hole seeding prescriptions based on gas properties, which produce $\sim 10^{4}$--$10^{5}\,M_{\odot}$ seeds in halos that contain star-forming, metal-poor gas.
\Abrahma\ combines the large cosmological volume and statistical power of \astrid\ with the physically motivated gas-based black hole seeding models from \brahma.
Motivated by JWST's discoveries of massive black holes (BHs) at $z\gtrsim 9$, \Abrahma\ adopts a lenient heavy-seed prescription from the \brahma\ suite, allowing for the formation of $4\times 10^{4-5}$~\Msun\ seeds in halos with star-forming, metal-poor gas.
The seeding model is motivated by scenarios in which heavy seeds form through stellar collisions in star clusters or from the rapid growth of Population III remnants.
The improved seeding model enables \Abrahma\ to form BH seeds much earlier and more efficiently compared to \astrid. 
%with the first seeds appearing at $z\sim 27$. 
This significantly enhances early BH growth, producing a $z=8$ BH number density more than an order of magnitude higher than that in \astrid\ over the mass range $10^{5-7}$~\Msun.
%However, rapid growth occurs only in a subset of environments: 
BHs reaching masses consistent with \gnz\ and \ceers\ typically originate in highly compact density peaks and undergo multiple early mergers.  
In these systems, $\sim50\%$ of BH masses by $z=11$ is from BH mergers, after which gas accretion becomes the dominant growth channel.
Without this early merger-driven assembly, \astrid\ cannot reproduce the high-mass BH detected by JWST. 
Our results indicate that abundant early seed formation combined with frequent mergers can explain several JWST massive BH candidates without requiring sustained super-Eddington accretion. 
As a testable prediction, \Abrahma\ yields $\approx4$ LISA detectable BH merger events per year at $z\geq8$, which is three orders of magnitude higher than that in \astrid.

\end{abstract}


\section{Introduction} \label{sec:intro}

In recent years, the James Webb Space Telescope (JWST) has rapidly transformed our view of massive black hole (BH) assembly in the early universe by unveiling a growing population of massive BHs at high redshift. 
While quasars at $z\gtrsim 6$ had already established that black holes with $M_{\rm BH}\gtrsim 10^{8-9}$~\Msun\ exist within the first Gyr \citep{Wang2021, Yang2020, Fan2023}, JWST is now identifying a much larger number density of low-luminosity AGN and BH candidates at similar redshifts \citep{Harikane2025, Matthee2024, Taylor2025}. 
Beyond increasing the sample size, JWST provides new constraints on the host environments of these high-$z$ massive BHs.
Based on JWST/NIRCam observations, the EIGER Collaboration \citep{Kashino2023} found a strong diversity among the observed quasar fields at $z\geq 6$ \citep{Eilers2024}. 
Using a compilation of JWST-identified low-luminosity AGNs at $5<z<6$, \citet{Arita2025} performed a clustering analysis and provided an
empirical estimate of their host dark matter (DM) halo mass: $\log M_{\rm halo}\approx 10^{11}\ h^{-1}M_{\odot}$. 
These discoveries suggest that early BH formation and growth may be both more common and more diverse than implied by the pre-JWST quasar census. 
Moreover, 
JWST is also pushing the redshift frontier for AGN by detecting a handful of objects at $z\sim 9-11$, such as \ceers\, \gnz, and UHZ1 \citep{Larson2023_ceers_obs, Maiolino2024_gnz11, Bogdan2024_uhz1_obs, Goulding2023_uhz1_obs}, which challenge our understanding of BH formation and growth theories.

Explaining the emergence of $\sim10^{6-7}$ \Msun black holes by $z=9$ is especially difficult~\citep{Volonteri2010, Johnson2016, Inayoshi2020}. Even under optimistic assumptions of continuous Eddington-limited accretion, the available time is short and requires a near-unity duty cycle. One possibility is to invoke sustained episodes of super-Eddington accretion \citep{Pezzulli2016, Inayoshi2016, Lupi2024}. However, feedback can substantially reduce the efficiency of such rapid growth \citep{Pacucci2015, Massonneau2023,2025A&A...704A.177P}. As a result, producing the most massive JWST-era black hole candidates through accretion alone remains extremely challenging.

Large-volume cosmological simulations are essential for studying this problem because they can self-consistently capture the complex interplay between BH growth and the evolving galaxy population across a wide range of environments. The \astrid\ simulation \citep{Bird2022,Ni2022_astrid, Zhou2025} was designed to provide statistically representative predictions for BH evolution in a large volume ($250\, h^{-1}{\rm Mpc}$ per side) across cosmic history. 
\astrid\ successfully reproduces a range of pre-JWST constraints on galaxies and BHs at high redshift \citep{Ni2022_astrid, Ni2025}, including the BH mass function, luminosity function, and $M_{\rm BH}-M_{\rm gal}$ scaling relations.
However, like many other cosmological simulations in the literature, \astrid\ adopts a BH seeding prescription based purely on a threshold halo mass.  This prescription cannot distinguish between different physical seeding channels. At the same time, \astrid\ does not produce the earliest, rapidly growing BH population implied by JWST detections, highlighting the need for improved BH formation models.

Several cosmological simulations have begun to implement more physically motivated prescriptions for BH seeding that depend on local gas conditions \citep{Taylor2014,Tremmel2017,Dubois2016_horionAGN,
Cenci2025,Bhowmick2024_brahma,Bhowmick2024_highz_SMBH}. In particular, the \brahma\ suite \citep{Bhowmick2022_gasspin_LW,
Bhowmick2022,
Bhowmick2024_brahma,Bhowmick2024_subgridmodel,Bhowmick2024_highz_SMBH} has systematically explored a wide range of gas-based seeding scenarios using prescriptions tied to properties such as gas density, metallicity, Lyman Werner~(LW) radiation, gas spin, and environmental richness. These studies demonstrate that the choice of seed model can strongly influence early BH growth and the resulting high-redshift BH populations observable with JWST. 
\citet{Bhowmick2025} used constrained \brahma\ cosmological simulations of rare ($5\sigma$) overdense peaks to study the assembly of the earliest $z\sim9$--11 BHs discovered by JWST under systematic variations of heavy seed models ($\sim10^{4}$--$10^{5}\,M_{\odot}$). They found that, under standard assumptions for stellar and AGN feedback, exceptionally high abundances of heavy seeds are required to assemble BHs with the current mass estimates of \gnz\ and \ceers. The required abundances are significantly larger than what is typically expected under canonical DCBH formation scenarios, which require extremely high LW fluxes ($\gtrsim1000\,J_{21}$; \citealt{Shang2010,Sugimura2014}). This is mainly because AGN feedback strongly suppresses early BH accretion in these \brahma\ simulations. With a large enough number of heavy $\sim10^5~M_{\odot}$ seeds, mergers can significantly accelerate BH growth at early times, which subsequently also boosts accretion at later times. In contrast, lower-mass seeds ($\lesssim10^{4}\,M_{\odot}$), even if formed in larger numbers, grow more slowly due to longer merger delay times resulting from weaker dynamical friction.

From a physical perspective, several promising pathways have been proposed that could produce heavy ($\sim10^{4}$--$10^{5}\,M_{\odot}$) seeds more efficiently. These include enhanced stellar and BH collisions in ultra-dense ($\gtrsim10^{8}\,\mathrm{cm^{-3}}$) nuclear star clusters (NSCs) \citep{Kritos2023,Pacucci2025}, supra-exponential BH accretion in NSCs \citep{Natarajan2021}, rapid BH mergers of $\sim10^{4}\,M_{\odot}$ seeds forming in feedback-free star clusters within early proto-galaxies via gravo-gyro instability \citep{Dekel2025}, hyper-Eddington growth of a small fraction of light Pop~III seeds soon after their formation \citep{Mehta2026}, or from the core collapse of self-interacting dark matter (SIDM) halos \citep{Feng2021, Jiang2026}. 
While these processes occur on scales far below the resolution of large cosmological simulations, their net outcome, namely the formation of $\sim10^{4}$--$10^{5}\,M_{\odot}$ seeds, can be incorporated in simulations such as \brahma\ by initializing BH seeds with these masses and following their subsequent growth through accretion and mergers.

A significant limitation of the constrained \brahma\ simulations of \citet{Bhowmick2025} is their relatively small volumes. While constraining the initial conditions allowed these simulations to target rare environments and explore a wide range of seed models at modest computational expense, it also makes it difficult to derive unbiased predictions for the number densities, environmental distributions, and clustering properties of high-redshift BHs. 
This limitation is particularly important given emerging evidence that the high-$z$ AGN population spans a wide range of environments \citep{Eilers2024}, requiring predictions that simultaneously capture both rare peaks and more typical overdense regions within a representative cosmological volume.

This leads us to develop \Abrahma\ (\textbf{A}STRID with \textbf{M}BH seeding from \textbf{BRA}MA), a new simulation that combines the statistical power and representative cosmological volume of \astrid\ with the physically motivated gas-based BH seeding prescriptions from \brahma. 
Based on the results of the constrained \brahma\ simulations, we adopt the most lenient seeding prescriptions allowed by our resolution limits, as these are the only models within the \brahma\ suite capable of producing a \gnz-like BH by $z\sim10$ \citep{Bhowmick2025}. 
\Abrahma\ uses the same volume, resolution, and initial condition realization as the original \astrid\ simulation, and retains the underlying \astrid\ galaxy formation model, modifying only the BH seeding model. 
Specifically, we remove the halo-mass threshold for seeding and allow seeds to form in all resolvable halos. 
Instead, seed formation occurs in halos where the gas is actively forming stars and is metal-poor. 
Our adopted seed masses range from $\sim4\times10^4$--$10^5\,M_{\odot}$, as in \astrid. 
Overall, this seed model choice can be regarded as an approximate upper limit among scenarios accessible at our resolution. Therefore, if any JWST BHs cannot be reproduced within \Abrahma, it would indicate additional seeding or growth channels that are beyond the scope of our current subgrid seeding and growth models. 

In this paper, we present the first results from \Abrahma\ evolved to $z=8$. 
This paper is organized as follows. 
Section~\ref{sec:method} describes the simulation setup, including the fiducial \astrid\ galaxy formation model and the modified BH seeding prescription. 
We present the results concerning the early growth of BH population in \Abrahma\ in Section~\ref{sec:results} and summarize our conclusions in Section~\ref{sec:conclusion}.

\section{Simulation Setup}
\label{sec:method}

\subsection{\astrid\ galaxy formation model}
%\aklantcomment{In this section, we can start with describing the MP GADGET code and general features of the ASTRID galaxy formation model, but don't refer to ASTRID BRAHMA quite yet}
%\aklantcomment{In this section, we show the analysis that is currently in the Appendix. We describe the BRAHMA seed models, and discuss how this behaves in concert with the ASTRID galaxy formation model. How the different choices we made in going from BRAHMA to ASTRID-BRAHMA, impact the seed formation history}
%\yihao{not sure how many details of ASTRID galaxy model we want to inlucde here, since we have described in detail in the ASTRID paper. Now I just highlight the star-formation model when we compare with BRAHMA, and provide the ASTRID paper as reference}

%In this section we briefly summarize the key features of the numerical setup and the \astrid\ galaxy formation model that are inherited by the \Abrahma\ simulations. 

Hosting $0.33$ trillion particles in a simulation box of $250\,{\rm Mpc}/h$ per side, ASTRID \citep{Bird2022,Ni2025, Zhou2025} is one of the largest hydrodynamical simulations evolved to $z=0$. 
It is performed using the massively parallel simulation code MP-GADGET \citep{Feng2018_mpgadget}, adopting a TreePM method for gravity calculation and a smoothed particle hydrodynamics (SPH) method for gas dynamics.
Its mass resolution is $m_{\rm DM}=6.74\times 10^{6}\,h^{-1}$~\Msun\ and $m_{\rm gas}=1.27\times 10^{6}\,h^{-1}$~\Msun\ in the initial conditions (ICs). The gravitational softening length is $\epsilon_{\rm g}=1.5\,h^{-1}$ kpc for both DM and gas particles. The ICs are set at $z=99$ and the cosmological parameters are from \citet{Planck}.
\Abrahma\ uses the same volume, resolution, and initial condition realization as the original \astrid\ simulation.

\astrid\ includes a full-physics sub-grid treatment for modeling galaxy formation, black holes, stellar and AGN feedback, and inhomogeneous reionization. 
\Abrahma\ inherits the same galaxy formation model as \astrid, and only modifies the BH seeding prescription.
In the following, we briefly summarize the key features the \astrid\ galaxy formation model and refer readers to \citet{Bird2022} and \citet{Ni2022_astrid} for more details.

In \astrid, gas is allowed to cool radiatively through primordial gas cooling following \citet{Katz1996}, as well as through metal-line cooling. The metal cooling rate is estimated by scaling a solar metallicity template according to the gas metallicity, following \citet{Wiersma2009}. 
Star formation is implemented based on the multi-phase star formation model in \citet{Springel2003}, where gas above a density threshold of $0.13\ {\rm cm}^{-3}$ is allowed to form stars.
A correction is included to account for the effects of molecular hydrogen on star formation at low metallicities, which is implemented according to \citet{Krumholz2011_h2cooling}.   
Stars are formed with one-fourth of the mass of gas particles. 
Type II supernova wind feedback is implemented following \citet{Okamoto2010}, assuming wind speeds proportional to the local DM velocity dispersion. 

\astrid\ models metal return by treating each star particle as a single stellar population with a Chabrier initial mass function (IMF) \citep{Chabrier2003}. 
We follow the general approach of \citet{Vogelsberger2013} and \citet{Pillepich2018}, while using our own mass and metal tables for AGB stars, Type Ia and Type II supernovae \citep{Karakas2010,Doherty2014_metallII,Doherty2014_metallIII, Kobayashi2006, Nomoto1997}.
To avoid excessive mass growth of gas particles, the mass returned to any gas particle is capped: any mass returned to a gas particle in excess of four times the initial gas mass is assumed to be retained within the star.

MBHs are evolved with subgrid prescriptions for seeding, accretion, feedback, and dynamics. 
We will describe the BH seeding prescriptions in detail in the next subsection, and here we briefly summarize the other aspects of the BH model. 
The BH accretion rate $\dot{M}_{\mathrm{BH}}$ is estimated using the Bondi-Hoyle formalism \citep{BondiHoyle1944, DiMatteo2005_BH_model}, which is based on the local properties of nearby gas particles:
\begin{equation}    
\label{equ:Mdot}
\dot{M}_{\mathrm{B}}=4\pi\alpha \,G^{2}\,M_{\mathrm{BH}}^{2}\,\rho_{\mathrm{BH}}\left(c_{\mathrm{s}}^{2}+v^{2}_{\mathrm{vel}}\right)^{-3/2},
\end{equation}
where $c_{\mathrm{s}}$ is the local sound speed, $\rho_{\mathrm{BH}}$ is the gas density around the BH, and $v_{\mathrm{vel}}$ is the velocity of the black hole relative to the surrounding gas. 
The dimensionless boost $\alpha=100$ is adopted to account for the underestimation of the accretion rate due to the unresolved interstellar medium. 
Super-Eddington accretion is allowed with an upper limit of twice the Eddington accretion rate $\dot{M}_{\mathrm{Edd}}$. Therefore, the black hole accretion rate $\dot{M}_{\mathrm{BH}}$ is determined by $\dot{M}_{\mathrm{BH}} = \min\left(\dot{M}_{\mathrm{B}}, 2\dot{M}_{\mathrm{Edd}}\right)$.

\astrid\ adopts two-mode AGN feedback: 
high accretion mode (or thermal feedback) and low accretion mode (or kinetic feedback). 
%The low-accretion mode is activated at $z=2.3$ and is implemented for BHs with $f_{\rm Edd}<\chi_{\rm thr}$. 
As the low-accretion mode only comes into play at low redshifts \citep[$z\lesssim 2$;][]{Weinberger2017_tng}, while not affecting the early growth of BHs, we describe only the high-accretion mode here.
%The high-accretion mode is applied to BHs with $f_{\rm Edd}\geq \chi_{\rm thr}$ for $z\leq 2.3$.
% The Eddington threshold $\chi_{\rm thr}$ is dependent on the BH mass: 
% \begin{equation}
%     \chi_{\rm thr} = \min \left[0.002\left(\frac{M_{\rm BH}}{M_{\rm crit}}\right)^{2},\  0.05\right],
%     \label{equ:chi_thr}
% \end{equation}
% where $M_{\rm crit}=5\times 10^{8}\,h^{-1}$~\Msun\ is the critical mass for AGN feedback. 
In this feedback mode, 5\% of the radiated energy $\Delta \dot{E}_{\rm high}=0.05\,\eta\, \dot{M}_{\rm BH}c^{2}$ is thermally injected into the gas within twice the radius of the SPH smoothing kernel of the BH. 
The energy injection is carried out according to the SPH kernel weight, without any preferred direction.

Following \citet{Tremmel2015} and \citet{Chen2022_DF}, a subgrid model is adopted 
to account for the unresolved dynamical friction for BHs from surrounding gas, stars, and dark matter. 
This model not only gives a better estimatation of the BH hardening timescale, but also provides well-defined BH trajectories and velocities, which allows us to impose a more physical criterion for BH mergers: mergers occur only for close, gravitationally bound pairs \citep{Bellovary2011,Tremmel2017}.
To alleviate artificial heating and stabilize the BH motion in the early growth phase, we assign a dynamical mass tracer, $M_{\rm dyn}=10^{7}\,h^{-1}$~\Msun\ (i.e., $1.5\times m_{\rm DM}$), which is used only for the gravity calculation. 
$M_{\rm dyn}$ is kept until $M_{\rm BH}$ grows above $M_{\rm dyn}$. After that, $M_{\rm BH}$ is used to calculate the dynamical friction force applied to the BHs.

%\subsection{Shortcomings of the BH seed model in fiducial \astrid\ }
\subsection{Fiducial BH seed model in ASTRID}
%For \Abrahma, we modify the BH seeding prescription while keeping the remaining physics identical to \astrid\ ~(as \aklant{summarized in the previous paragraph})

The fiducial \astrid\ simulation adopts a halo-based BH seeding prescription.
Halos are identified using a friends-of-friends (FOF) algorithm \citep{Davis1985} with a linking length of 0.2 times the mean particle separation, and requiring each halo to contain at least 32 dark matter particles.
A halo is eligible for seeding if it satisfies both a total halo mass of $M_{\rm halo,FOF} > 5 \times 10^9~h^{-1}\rm{M_\odot}$ and a stellar mass of $M_{\rm *,FOF} > 2 \times 10^6~h^{-1}\rm{M_\odot}$.
The stellar-mass threshold ensures that BHs are seeded only in halos that contain sufficient cold, dense gas to form stars.
The BH seed mass $M_{\rm seed}$ is stochastically drawn from a power-law distribution of heavy seed masses ranging between $3\times10^{4}~h^{-1}\rm{M_\odot}$ and $3\times10^{5}~h^{-1}\rm{M_\odot}$.

A limitation of the \astrid\ seed model is the relatively high halo-mass threshold adopted for BH seeding.
While similar thresholds have been used in several other cosmological simulations~(e.g., $M_{\rm halo,FOF}={7.4}\times 10^{10}$~\Msun\ for TNG100, $1.48\times 10^{10}$~\Msun\ for EAGLE and $10^{9.5}$~\Msun\ for SIMBA),
star formation is known to occur in halos significantly below this mass scale.
Because of this, the onset of BH seeding may be artificially delayed relative to the physical channels expected to produce these heavy seeds, such as rapidly growing Pop III remnants, remnants of runaway stellar mergers or gas accretion in NSCs, or direct collapse black holes (DCBHs).
At the same time, we also find that the \astrid\ simulation does not form and grow BHs early enough to reproduce some of the most massive high-redshift BHs recently discovered by JWST.
Finally, we note that the \astrid\ seed model lacks any metallicity dependence. This is in contrast to most physical seeding channels that are expected to operate primarily in low-metallicity environments. 
Omitting the anticipated metallicity dependence of the seeding model could lead to the overproduction of seeds at low redshifts. 
%Considering the complex astrophysical process involving BH seed formation, we allow halos with the same mass to host different BH seeds. 

\begin{figure*}[!t]
    \centering
    \includegraphics[width=1\linewidth]{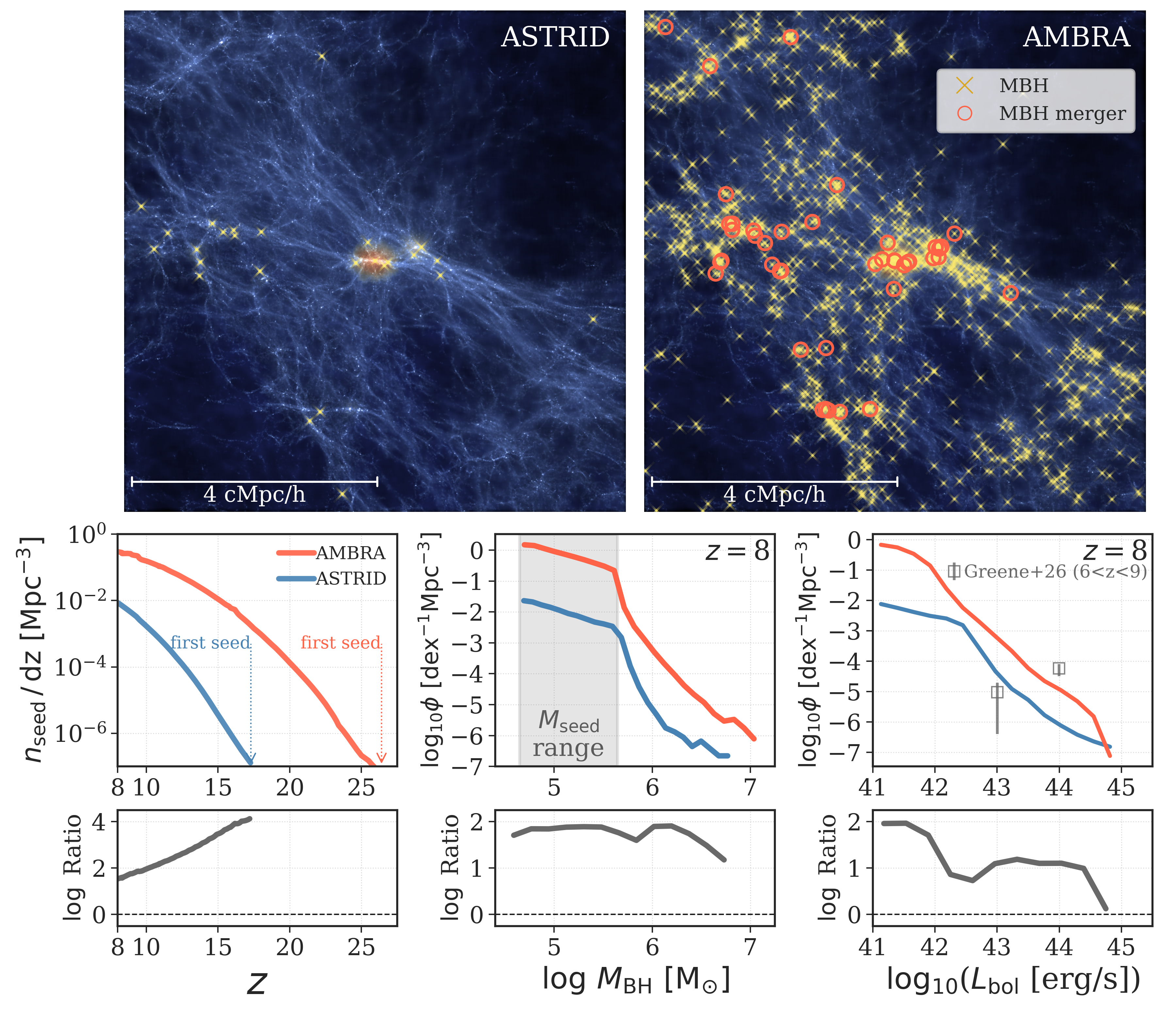}
    \caption{%\aklantcomment{Could we split the visualizations and the abundance plots into two separate figures with their own captions. You can still keep them within a common ``includegraphic" so they appear together, but splitting the captions will make it easier to refer to the different panels within the text} Comparison between \Abrahma\ and \astrid\ at $z=8$. 
    \textit{Top row}: large-scale environment of the most massive BH at $z=8$ in \Abrahma\ (right) and the same region in \astrid\ (left). 
    The visualization shows the gas density field in a box of $8\ {\rm cMpc/h}$ per side colored by temperature, from red to blue, indicating warm to cold. The same colorbar is applied to both panels. 
    The yellow crosses mark all the BHs with $M_{\rm BH}\geq10^{5}$~\Msun, and the red circles mark the remnants of mergers that occur during $8<z\leq9$ (there is no such merger in the plotted \astrid\ region).  
    %Throughout the panels in the middle row, red curves show \Abrahma\ and blue curves show \astrid. 
    \textit{Middle row:} from left to right, we show the BH seed formation history, the BH mass function, and the BH luminosity function at $z=8$.
    In all three panels, red curves show \Abrahma\ and blue curves show \astrid. 
    In the left panel, arrows mark the redshift of the first seed formation: $z=26.4$ for \Abrahma\ and $z=17.3$ for \astrid.
    In the middle panel, the gray band represents the adopted seed mass range ($3\times 10^{4}\leq M_{\rm seed}\leq 3\times 10^{5}\ h^{-1}$~\Msun).
    In the right panel, the luminosity function is compared with the observational constraints from \citet{Greene2026}.
    \textit{Bottom row}: the ratio of the quantities predicted by the two simulation~(\Abrahma/\astrid) for the corresponding panels above.
    The dashed horizontal line marks a ratio of 1.
    Overall, these comparisons demonstrate that \Abrahma\ seeds BHs more efficiently, and produces a larger population of massive BHs at high redshift than \astrid.
    }\label{fig:bhpopulation_compare}
\end{figure*}

\subsection{The \Abrahma\ seed model}

In the heavy-seed models explored by the \brahma\ simulations of \cite{Bhowmick2025}, seed formation is allowed in all sufficiently resolved halos (i.e., those containing $>32$ DM particles) provided that they satisfy critical thresholds in dense, metal-poor gas mass and Lyman-Werner (LW) radiation flux. 
Specifically, seeding occurs in halos containing sufficient dense ($\geq 0.13~\rm cm^{-3}$), metal-poor ($\leq10^{-4}~Z_{\odot}$) gas and are exposed to LW fluxes of $\sim10$--$300~J_{21}$. 
This framework addresses both limitations of the \astrid\ seed model: 
(1) it explicitly restricts seeding to low-metallicity environments, and (2) it does not introduce additional halo-mass threshold beyond that imposed by the simulation mass resolution.
Among the models explored by \cite{Bhowmick2025}, only the most lenient variants that require the smallest amounts of dense, metal-poor gas ($\sim5~M_{\rm seed}$) and little or no LW radiation ($\lesssim10~J_{21}$) were able to reproduce the mass of \gnz. 
Motivated by this result, we adopt this dense, metal-poor gas mass criterion in \Abrahma. 

However, we exclude the LW flux criterion in our implementation. Critical LW fluxes as low as $\sim10~J_{21}$ can induce DCBH formation only if additional processes, such as dynamical heating, suppress $\mathrm{H}_2$ cooling \citep{Wise2019}. These additional requirements can make the model significantly more restrictive~\citep{Bhowmick2024_highz_SMBH}. Instead, we adopt a heavy-seed formation scenario in which star formation within dense, metal-poor gas does not need to be suppressed by external LW radiation.
This choice is more aligned with seeds forming through runaway stellar collisions in nuclear star clusters (NSCs)~\citep{Kritos2023}, or through the end states of rapid hyper-Eddington growth of Population III remnants~\citep{Mehta2026}.

Overall, the \Abrahma\ seed model comprises a single primary seeding criterion: a minimum threshold in star-forming, metal-poor gas mass, hereafter denoted as $M_{\rm sfmp}$. In addition, there is an implicit halo-mass threshold~($M_{\rm h}$) set by the simulation mass resolution, corresponding to $M_{\rm h}=2\times10^{8}\,h^{-1}$ \Msun. 
This is $\sim25$ times smaller than the halo-mass threshold imposed in the original \astrid\ simulation, leading to a substantially higher abundance of seeds. 
Finally, BH seed masses are still drawn from the same power-law distribution adopted in \astrid, spanning from $\times10^{4}~h^{-1}\rm{M_\odot}$ to $3\times10^{5}~h^{-1}\rm{M_\odot}$.

Notably, the heavy-seed based \brahma\ simulations from \cite{Bhowmick2024_highz_SMBH} were run at $\sim8$ times higher resolution than the original \astrid. Achieving such resolution at the \astrid\ volume would require unprecedented particle counts of $\sim2\times11000^3$. 
To keep our computational expense feasible and to ensure an even-handed comparison with the original \astrid, we adopt the same resolution, volume, and initial-condition realization as \astrid. To select the most lenient seed model possible, we choose the smallest value of $M_{\rm sfmp}$ allowed at this resolution, corresponding essentially to a single star-forming metal-poor gas particle. This amounts to $M_{\rm sfmp}=1.27\times10^6\ ~h^{-1}~M_{\odot}$. We perform a resolution test in Appendix~\ref{app:resolution}, which demonstrates that running this simulation at the \brahma\ resolution would only lead to a slight deviation (by a factor of $\sim2$) in the seed formation history. This difference does not affect the main conclusions of this work. 

\begin{figure*}[!t]
    \centering
    \includegraphics[width=0.75\linewidth]{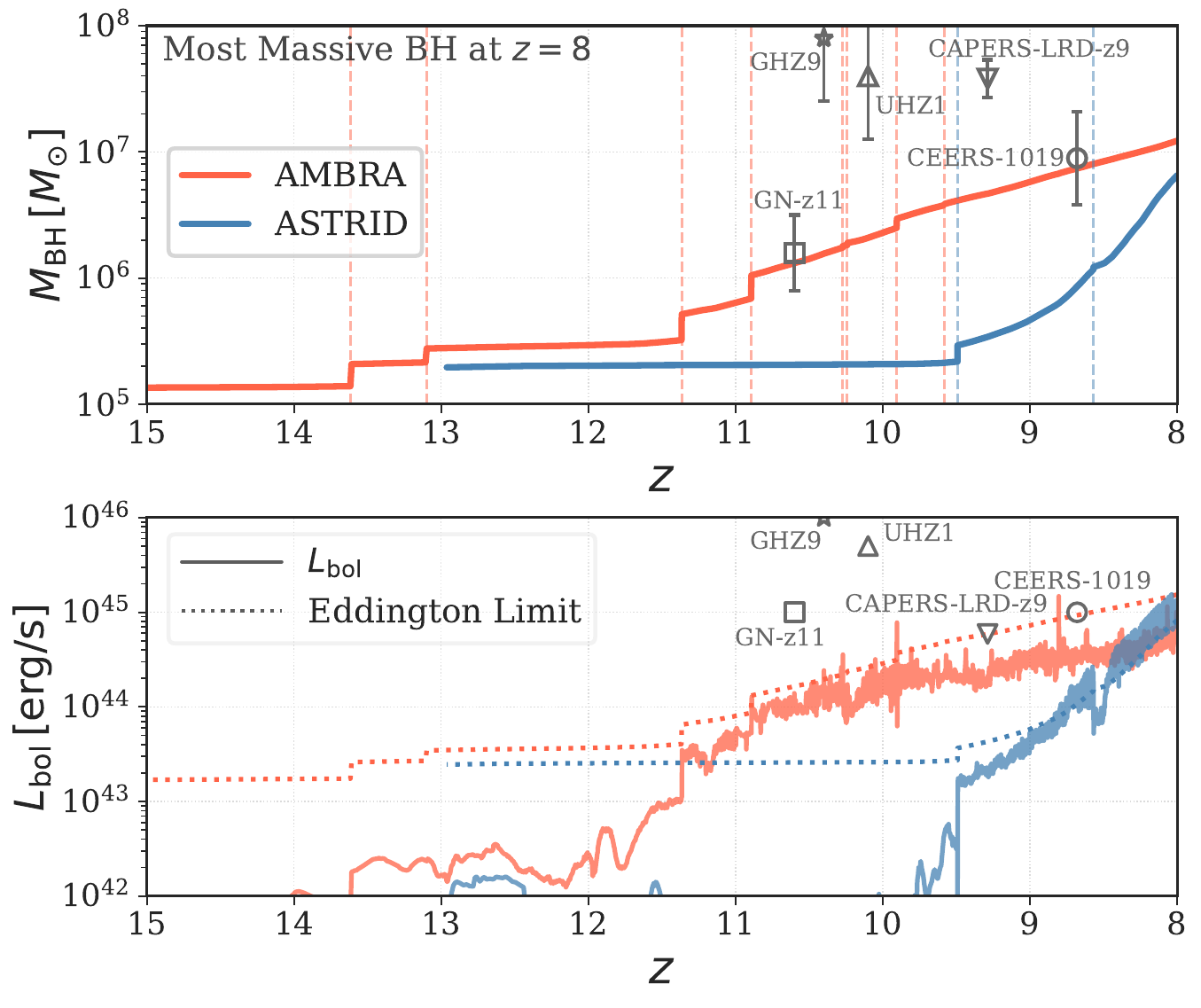}
    \caption{The evolution of the most massive BH at $z=8$ in \Abrahma\ (blue) and \astrid\ (red).
    Gray points with error bars show observed high-$z$ massive BHs population: \ceers\ \citep{Larson2023_ceers_obs}, 
    UHZ1 \citep{Bogdan2024_uhz1_obs, Goulding2023_uhz1_obs}, 
    \gnz\ \citep{Maiolino2024_gnz11, Tacchella2023_gnz11_obs},
    CAPERS-LRD-z9 \citep{Taylor2025_capers_LRD_z9_obs},
    and GHZ9 \citep{Kovacs2024_ghz9_obs}.  
    Among these observed BHs,
    two of which (\gnz\ and \ceers) are reproduced by \Abrahma. 
    \textit{Top:} Mass evolution of the most massive progenitors in both simulations.
    The vertical dash lines mark the merger events.
        \textit{Bottom:} Bolometric luminosity history (solid) of these progenitors. The dotted curves show the corresponding Eddington limits.
        %\aklantcomment{Add the observed luminosities here too}. 
    Compared to \astrid, the most massive BH in \Abrahma\ grows much more rapidly at $z\gtrsim 10$, and reaches a mass of $\sim10^{7}$~\Msun\ at $z=8$, which is about an order of magnitude higher than the most massive BH in \astrid.
  }\label{fig:massive_bh_compare}
\end{figure*}

\begin{figure*}[!t]
    \centering
    \includegraphics[width=0.85\linewidth]{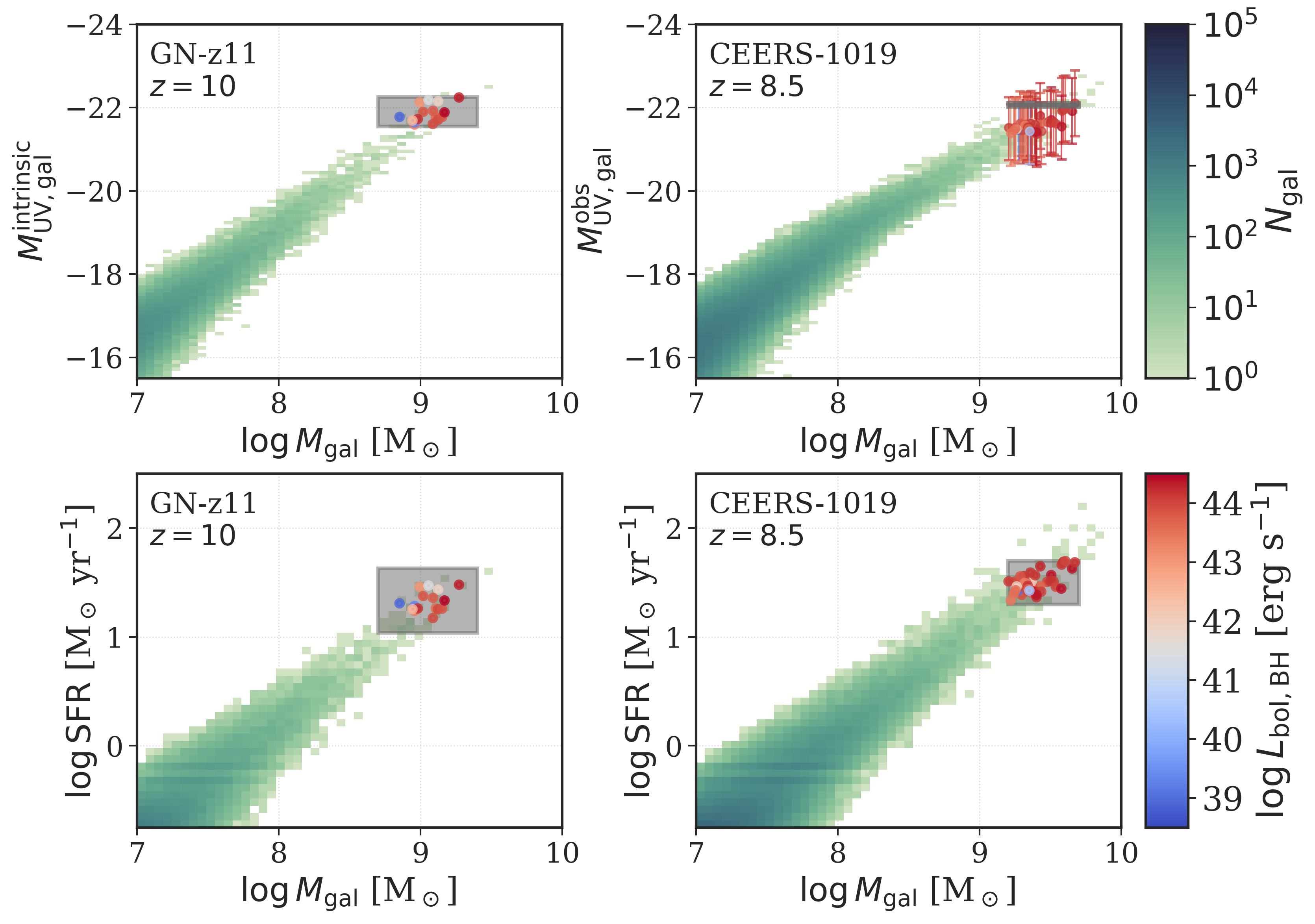}
    \caption{The counterparts of \gnz\ and \ceers\ (the dots) compared to the entire galaxy population in \Abrahma\ (the green pixels).
        The left column is plotted based on \Abrahma\ $z=10$ data, where we search for the counterpart for \gnz; and the right column is based on \Abrahma\ $z=8.5$ data, where we search for the counterpart for \ceers.
    \textit{Upper:} the UV magnitude of the galaxy $M_{\rm UV, gal}$ versus the galaxy mass $M_{\rm gal}$.
    \textit{Lower:} the galaxy SFR versus $M_{\rm gal}$.
    In each panel, the gray area corresponds to the observational constraints for GN-z11 and CEERS-1019 from \citet{Tacchella2023_gnz11_obs} and \citet{Larson2023_ceers_obs}, respectively. 
    We use the galaxy properties, including $M_{\rm UV}$, $M_{\rm gal}$, and ${\rm SFR}$, to identify the counterparts, and color these counterparts based on the $L_{\rm bol}$ of their central BH (the lower color bar). We hereafter study the BH properties hosted by these counterparts and compare with the JWST-based measurements.
    }\label{fig:galaxy_scatter}
\end{figure*}

\section{Results for ASTRID-BRAHMA Simulation} \label{sec:results}

%\aklantcomment{Finally after showing the analysis of the previous section, we can introduce ASTRID-BRAHMA using Figure 1}

\subsection{BH population at $z=8$}
% \aklantcomment{I'm editing this subsection to remove ``figure caption-like" language and focus more on the science takeaways. This helps making the text more concise and interesting to read imo} 

% \aklantcomment{Could you take this as an example and try to implement in the remaining subsections as well? More specifically, I'd try to minimize statements that solely refer to figure details~(solid lines, dashed lines, colors) etc. Those are already described in the figure captions, so they are redundant in the main text and makes it verbose.}

Fig.~\ref{fig:bhpopulation_compare} compares the BH populations produced by \Abrahma\ and \astrid\ at $z=8$. The top right panel shows the large-scale environments surrounding the most massive BH in \Abrahma\ at this redshift, and the top left panel shows the same region in \astrid. 
While both simulations produce similar large-scale structures, \Abrahma\ exhibits an overall BH number density that is approximately $\approx 55$ times higher than that in \astrid. 
This dramatic increase in BH abundance also leads to a significant number of BH mergers in this region in \Abrahma\ (red circles in the top-right panel). In contrast, \astrid\ produces no merger events in the same region at $z\gtrsim8$.

In the middle row of Fig.~\ref{fig:bhpopulation_compare}, we plot the BH seed number density as a function of redshift (left), the BH mass function (middle), and the luminosity function (right) at $z=8$; and we show the logarithmic ratio between these two simulations in the bottom row.
With the improved seeding model, \Abrahma\ seeds BHs at a much earlier time.
The first seed forms at $z=27$ in \Abrahma, and the seed number density reaches $0.01/{dz}/{\rm Mpc}^{3}$ at $z\approx 15$. In \astrid, the first seed does not appear until $z=17$, and the seed number density reaches $0.01/{dz}/{\rm Mpc}^{3}$ at $z\approx 8$.

The BH mass functions in \Abrahma\ maintain a higher normalization across the entire BH mass range probed by the simulation volume ($3\times10^4 \sim 10^7~M_{\odot}$). 
The slopes of the mass functions naturally steepen for both simulations at $>3\times10^5~M_{\odot}$, i.e., beyond the range of the initial seed masses.
This occurs because, at this early phase, accretion-driven BH growth remains weak in both simulations. The slopes become less steep at higher BH masses as accretion-driven growth becomes more significant. 
Interestingly, at the high mass end the gap between \Abrahma\ and \astrid\ narrows: the ratio between the two mass functions decreases from $\sim100$ at $M_{\rm BH}\sim 10^{6}$~\Msun\ to $\sim10$ at $M_{\rm BH}\sim 10^{7}$~\Msun.
This is also evident at the brightest end ($L_{\rm bol}>44$~erg/s) of the AGN luminosity function (middle right panel). 
Specifically, \Abrahma\ produces a higher normalization of the AGN luminosity function over most of the luminosity range probed by our volume ($\sim10^{41}$--$10^{45}~\rm erg~s^{-1}$).
Compared to \astrid, \Abrahma\ produces more consistent results with JWST detection (gray dots) \citep{Greene2026} at $L\sim10^{44}$~erg/s. 
However, this trend reverses at the brightest end, where \astrid\ has a higher normalization.
As we shall see in the next section, this reflects stronger late-stage accretion (occurring at $z\lesssim 9$) in \astrid\ compared to \Abrahma.
%\aklantcomment{Just realized that we don't really have any text about the AGN LF comparison with Greene et al results in the main text. We should add a statement here}

% (\aklantcomment{I think if we also show ratio plots, the above trends would be much more visible and interesting to point out, as they come up later in Figure 3 too. It might also be interesting to show the Eddington ratio functions.})

%\aklantcomment{Lastly, I think it is also worth pointing out the difference in the faint end slope of the AGN LFs between the two simulations. ASTRID produces a sudden flattening at the faint end which is not seen in ASTRID-BRAHMA. I bet this is because the lowest mass halos are not populated in ASTRID. It would therefore be interesting to show (in red dashed line), the ASTRID BRAHMA LFs but only for halos above the seeding threshold for ASTRID. My guess is that this would produce a similar flattening at the faint end as ASTRID.}

\subsection{The most massive BH}

Having examined the full population of BHs produced by \Abrahma\ at $z=8$, we now focus on the most massive BHs and compare them with the high-$z$ BHs discovered by JWST. Fig.~\ref{fig:massive_bh_compare} shows the evolutionary history of the most massive BH in \Abrahma\ (red) and in \astrid\ (blue) by $z=8$, obtained by tracing their most massive progenitor branch through each merger event.
To guide the eye, we plot the observed high-$z$ massive BHs population as gray points with error bars, including \ceers\ \citep{Larson2023_ceers_obs}, UHZ1 \citep{Bogdan2024_uhz1_obs, Goulding2023_uhz1_obs}, \gnz\ \citep{Maiolino2024_gnz11, Tacchella2023_gnz11_obs}, CAPERS-LRD-z9 \citep{Taylor2025_capers_LRD_z9_obs}, and GHZ9 \citep{Kovacs2024_ghz9_obs}.
For CAPERS-LRD-z9, we estimate its bolometric luminosity from the observed broad H$\beta$ line, $F_{{\rm H}\beta,{\rm board}}$, reported in \citet{Taylor2025_capers_LRD_z9_obs}.
We first convert the broad H$\beta$ flux to H$\alpha$ assuming the intrinsic Case B ratio $H_{\alpha}/H_{\beta}$ = 2.86 \citep{Hummer1987}.
We then apply the bolometric correction from \citet{Stern2012}, $L_{\rm bol}=130 L_{{\rm H}\alpha,{\rm broad}}$, which yields $L_{\rm bol} = 6\times 10^{44}$~erg/s.

At $z=8$, the most massive BH in \Abrahma\ reaches $\sim1.2\times10^{7}~M_{\odot}$.
Its progenitor seed forms at $z\sim21.7$ with a mass of $\sim10^5~M_{\odot}$. 
Accretion-driven BH growth is relatively weak at these earliest stages, with accretion rates (or luminosities) $\lesssim10$ times below the Eddington rate at $z\geq 11$, as shown in the bottom panel. 
Four merger events boost the seed mass to $\sim10^{6}~M_{\odot}$ by $z\sim11$, making the $M_{\rm BH}$ consistent with the measured mass of \gnz \footnote{A caveat here is that the most massive BH at $z=8$ is not the most massive object in the simulation box at $z=11$, but their mass difference at $z=11$ is small (within a factor of two) and does not affect our conclusion.}.
Since $z=11$, the gas accretion is significantly enhanced, with the overall $L_{\rm bol}$ close to the Eddington limit along with occasional spikes into the super-Eddington regime. 
The estimated bolometric luminosity of \gnz\ is still higher than these super-Eddington spikes; however, note that \gnz\ is reported to have an Eddington ratio of $\sim5$, whereas we cap our accretion rates at $\sim2\times$ Eddington. 
At $z\lesssim11$, the BH continues to grow steadily under mildly sub-Eddington accretion and reaches a mass of $\sim6\times10^6~M_{\odot}$ by $z\sim8.7$, consistent with the measured BH mass of \ceers. 
By this time, the accretion rates become more sub-Eddington ($\sim5$--$6$ times below Eddington), likely due to AGN feedback. 
The observed bolometric luminosity of \ceers\ lies above these typical accretion rates, but remains consistent with the peaks of accretion rates given the strong variability. 
%\aklantcomment{Again, I'm basing this statement on Figure 5, but we should add the observed luminosities to Figure 3 as well!}
 
%within the \astrid\ simulation (blue cueves in Fig.~\ref{fig:massive_bh_compare}), 
In \astrid, the growth of the most massive BH (blue curves in Fig.~\ref{fig:mass_contribution}) is substantially delayed compared to \Abrahma. The BH forms at $z\sim13$ with a fairly high initial seed mass of $2\times10^5~M_{\odot}$, but undergoes little growth until the first merger occurs at $z\sim9.5$. The BH finally begins to accrete rapidly at $z\lesssim9$, with accretion rates consistently above the Eddington limit. By $z\sim8$, the luminosities produced by the \astrid\ BH exceed those of \Abrahma, as we also noted at the brightest end of the $z=8$ luminosity functions in the previous section.
This likely occurs because the lack of earlier BH accretion episodes in \astrid\ allows gas to accumulate to higher densities, leading to more rapid growth at later times. Nevertheless, this late-time rapid accretion is not sufficient for the BH to catch up with \Abrahma\ by $z=8$. The resulting BH masses for \astrid~ remain $\sim10$ times below the \ceers\ and \gnz\ measurements at their respective redshifts. 
At $z=8$, the most massive BH in \astrid\ is $\sim6\times10^6~M_{\odot}$, $\sim2$ times smaller than that of \Abrahma\ .

%observed high-$z$ BH masses considered here. 
%At $z=8.5$, the detected redshift of \ceers, the most massive BH is only %$M_{\rm BH}\approx10^{6}$~\Msun. 

Finally, we note that despite the enhanced BH growth in \Abrahma\ compared to \astrid, there are no BHs that come close to the estimated masses of UHZ1, GHZ9, and CAPERS-LRD-z9, which are over $10^{7}$~\Msun\ at $z>9$. 
With that being said, there are substantial uncertainties in the high-$z$ BH mass estimates in general. 
For example, the UHZ1 estimate is based on the X-ray luminosity under the assumption of Eddington growth \citep{Bogdan2024_uhz1_obs}. 
A recent reanalysis of UHZ1 finds that the original reported hard X-ray excess is not robustly reproducible, while independent JWST spectroscopy reveals no clear AGN signatures \citep{Zou2026, Alvarez-Marquez2026}. 
%However, it could also correspond to a lower-mass BH undergoing super-Eddington growth. 
The BH mass estimates for LRDs such as CAPERS-LRD-z9 could also be overestimated \citep{Naidu2025,Rusakov2026}. 
The existence of such extraordinarily massive BHs may require other BH formation and growth scenarios that are not captured within our simulations. Possible channels could be accelerated accretion-driven growth under weaker AGN/stellar feedback compared to what is required to reproduce low-$z$ galaxies and BHs \citep{Bhowmick2025}, or primordial BHs~\citep{Ziparo2025, Zhang2025}, core-collapse of SIDM halos \citep{Feng2021}, and ultramassive seeds \citep{Chon2025, Mayer2024}.

% This includes two possibilities to boost BH growth at high $z$ explored by \citet{Bhowmick2025}: (1) rapid merger-driven BH growth of heavy seeds compared to our subgrid-dynamical-friction model predictions, and (2) accelerated accretion-driven growth under weaker AGN/stellar feedback compared to what is required to reproduce low-$z$ galaxies and BHs. Other possibilities include primordial BHs~\citep{Ziparo2025, Zhang2025}, as well as ultramassive seeds \citep{Chon2025, Mayer2024}.

%BH mass grows through gas accretion or mergers. In the following, we investigate the contribution of these two channels in the growth of the two BHs plotted in Fig.~\ref{fig:massive_bh_compare}.
Overall, we find that in \Abrahma, mergers play an important role in accelerating BH growth at the earliest stages, and subsequently enhance gas accretion, which is proportional to $M_{\rm BH}^{2}$.
This enables \Abrahma\ to reproduce BH masses of $\sim10^6$--$10^7~M_{\odot}$ by $z\sim9$--$11$, consistent with the inferred masses of 
\gnz\ and \ceers.
Their observed luminosities can correspond to a phase near the peak of the instantaneous accretion.
In contrast, without these early mergers, \astrid\ fails to reproduce either the masses or the luminosities of these two JWST BHs.

\subsection{Simulation counterparts of \gnz\ and \ceers}\label{sec:counterpart}

\subsubsection{Identifying galaxy counterparts}

In the previous section, we showed that \Abrahma\ does produce BHs with masses comparable to two observed high-$z$ objects: \gnz\ and \ceers. 
%In the following, we perform a more detailed search for simulated counterparts to these two objects.
However, both of these objects were first identified as galaxies \citep{Oesch2014, Oesch2016, Zitrin2015_ceers1019, Roberts-Borsani2016_ceers1019}, with their central AGN revealed through follow-up observations \citep{Tacchella2023_gnz11_obs, Larson2023_ceers_obs}. This motivates us to search for their counterparts in \Abrahma\ using only host-galaxy properties, and to explore the general BH population hosted by these counterparts. In the process, we shall investigate whether the observed BHs in \gnz\ and \ceers\ correspond to typical populations within these galaxies, or whether they preferentially emerge in a small subset of these galaxies depending on their environment.

\begin{figure*}[!t]
    \includegraphics[width=\linewidth]{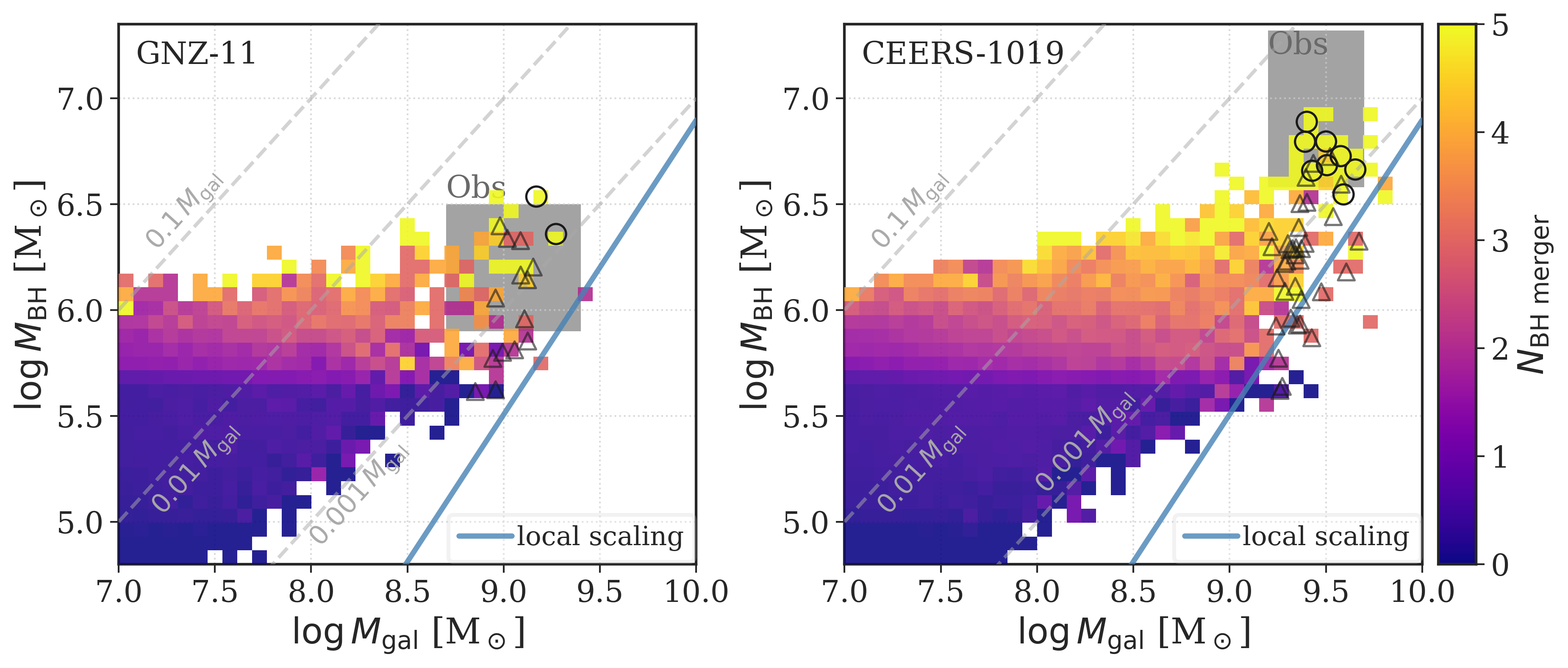}
    \caption{
    $M_{\rm BH}$--$M_{\rm gal}$ scaling relation for the central BH population at $z=10$ (left column) and $z=8.5$ (right column) in \Abrahma, where we search for \gnz\ and \ceers\ counterparts, respectively. The black markers are the central BH of the counterparts, and the underlying pixels are the distribution of the central BH population at the corresponding redshifts.
    % The scaling relation between the central BH $M_{\rm BH}$ and $M_{\rm gal}$. 
    We color the distribution based on the number of mergers 
    the BHs experienced. We count only the mergers of the 
    massive progenitors. 
    The color is averaged among the objects in each pixel, and the two panels share the same color scale according to the colorbar plotted on the right. 
    To guide the eye, we plot the level of $M_{\rm BH} = 0.001,\, 0.01, 0.1\ M_{\rm gal}$ using the gray dashed lines, and the local scaling relation from \citet{Greene2020} using the blue line.
    The gray area represents the observational constraints for \gnz\ and \ceers. 
    Same as Fig.~\ref{fig:Lbol_Mbh}, for the BHs hosted by the galaxy counterparts, circles represent those whose peak luminosity within $50$~Myr is above 50\% of the observed $L_{\rm bol}$, and others are plotted by triangles.   
    }\label{fig:Mbh_Mstar} 
\end{figure*}

\begin{figure*}[!t]
    \centering
    \includegraphics[width=0.9\linewidth]{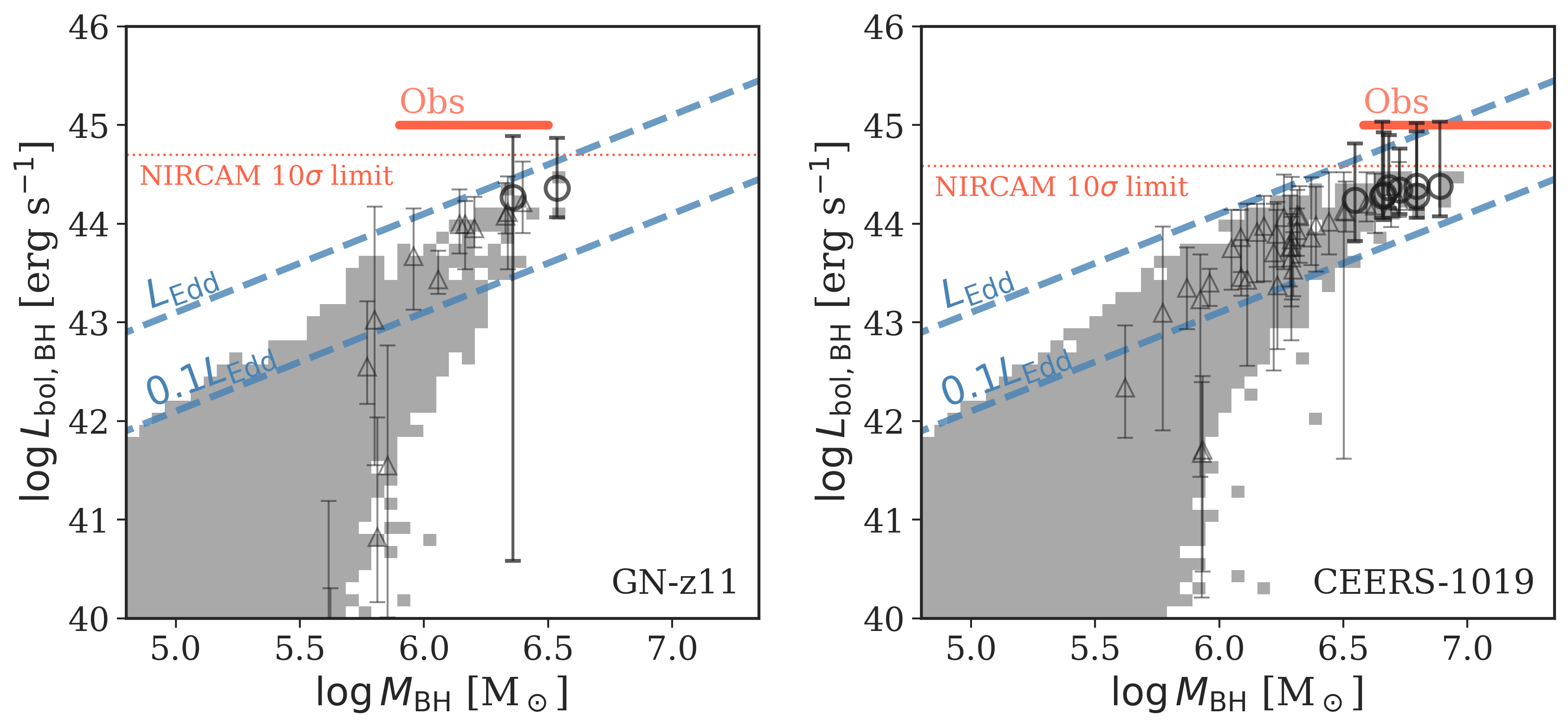}
    \caption{$L_{\rm bol}$--$M_{\rm BH}$ scaling relation for the central BH population at $z=10$ (left column) and $z=8.5$ (right column) in \Abrahma, where we search for \gnz\ and \ceers\ counterparts, respectively. The black markers are the central BH of the counterparts, and the underlying pixels are the distribution of the central BH population at the corresponding redshifts.
    % \textit{Lower:} the BH bolometric luminosity $L_{\rm bol}$ versus the BH mass $M_{\rm BH}$. 
    The blue dashed lines mark one and 10\% of the Eddington limit, and the red bar labels the observational constraints.  
    We also plot the detection limits of JWST-NIRCam (red dotted) derived using bolometric corrections from \citet{Shen2020}.
    To indicate the rapid fluctuations of BH luminosity, we show the $L_{\rm bol}$ range within 50 Myr using the error bars in the bottom panels.
    Circles highlight the BHs whose peak $L_{\rm bol}$ during this time is above $5\times 10^{44}$ erg/s (i.e., 50\% of the observed $L_{\rm bol}$), and other objects are marked by triangles. 
    For galaxies counterparts, their central BH mass can vary by more than two orders of magnitude even though they have similar $M_{\rm gal}$, and the BHs with more mergers tend to be more massive.
    The observed BHs select the brightest objects among the population at the given redshift, which are close to or even above the Eddington limit. 
    }\label{fig:Lbol_Mbh}
\end{figure*}

To identify galaxy counterparts, we match three observables to the available constraints: stellar mass $M_{\rm gal}$, star-formation rate (SFR), and rest-frame UV magnitude $M_{\rm UV,gal}$. 
The SFR for \Abrahma\ galaxy population is derived from the stellar mass formed over the last 10 Myr, making it broadly comparable to observational estimates based on the H$\alpha$ star-formation tracer.
To calculate the intrinsic UV magnitude of galaxies $M_{\rm UV, gal}^{\rm intrinsic}$, we construct their spectral energy distributions. We model each star particle as a simple stellar population (SSP) with its birth time, metallicity, and mass extracted from the simulation. We use the \textsc{FSPS} stellar population synthesis code \citep{Conroy2009, Conroy2010} with the PARSEC isochrones \citep{Bressan2012_parsec} and MILES stellar library \citep{Sanchez-Blazquez2006_miles}, assuming a Chabrier initial mass function \citep{Chabrier2003}. 
The luminosity for an individual galaxy is the sum of the emission of all star particles in this galaxy. 
We match the intrinsic magnitude $M_{\rm UV, gal}^{\rm intrinsic}$ for \gnz, which is provided by \citet{Tacchella2023_gnz11_obs}.
For \ceers, only the observed UV magnitude is available in \citet{Larson2023_ceers_obs}, so we convert $M_{\rm UV, gal}^{\rm intrinsic}$ to observed magnitude $M_{\rm UV,gal}^{\rm obs}$ by applying the dust attenuation model according to Model A in \citet{Vogelsberger2020}. This is an empirical scaling relation to link dust-free rest-frame UV magnitudes with observed dust-attenuated rest-frame UV magnitudes.

% When only the  only the observed magnitude is available, we convert $M_{\rm UV, gal}^{\rm intrinsic}$ to observed magnitude $M_{\rm UV,gal}^{\rm obs}$ by applying the dust attenuation model according to Model A in \citet{Vogelsberger2020}. This is an empirical scaling relation to link dust-free rest-frame UV magnitudes with observed dust-attenuated rest-frame UV magnitudes \aklantcomment{The last part of this para, starting from "We match.." is not very clear. Could you please rephrase?}.

\begin{figure*}[!t]
    \centering
    \includegraphics[width=0.9\linewidth]{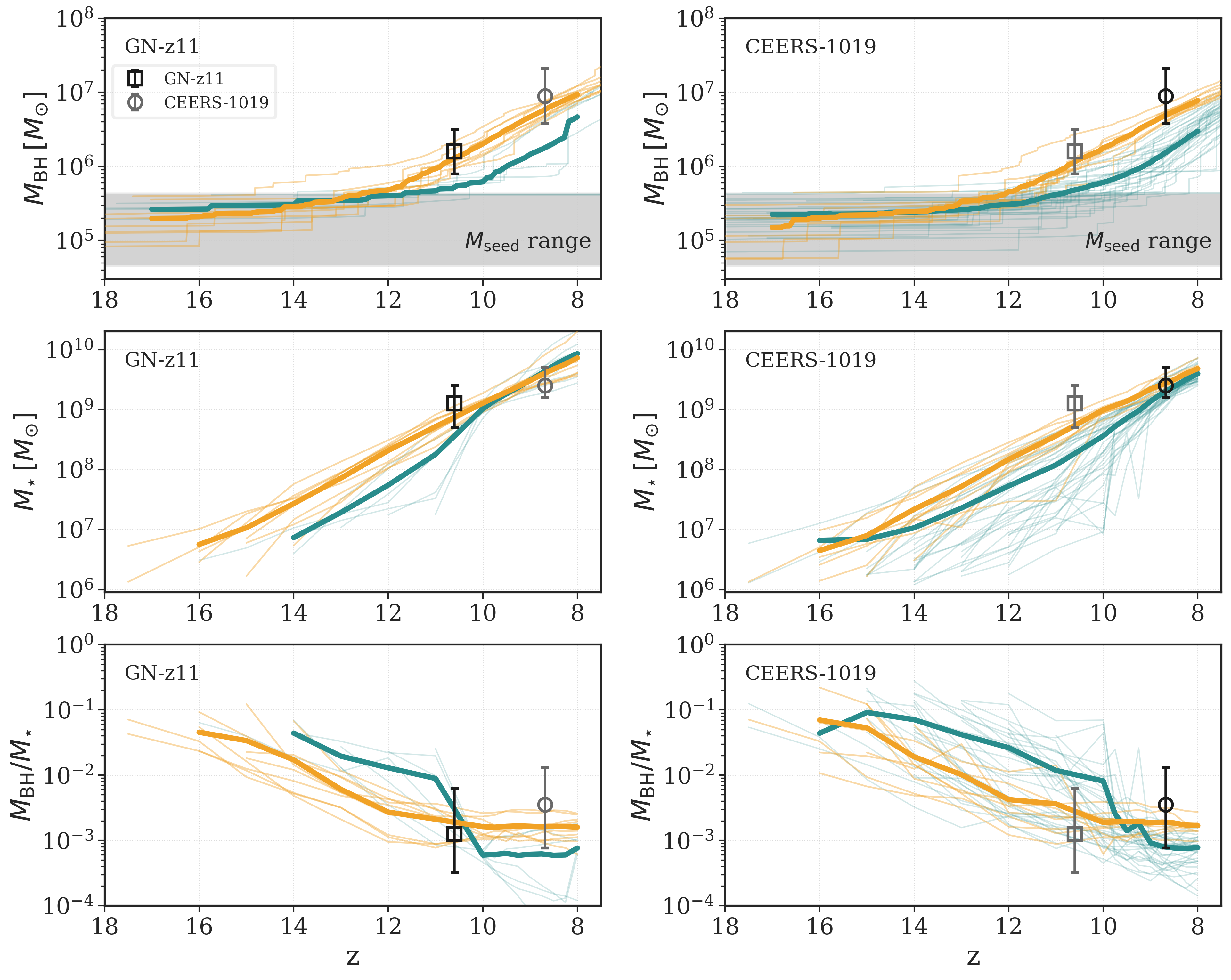}
    \caption{Evolution histories of the BHs within the \gnz\ (left column) and \ceers\ (right column) counterparts.    
    From top to bottom, we show the evolution of central BH mass $M_{\rm BH}$, the galaxy mass $M_{\rm gal}$, and the mass ratio $M_{\rm BH}/M_{\rm gal}$. 
    In each panel, the black error bar marks the observational constraint for the target objects in that column, while the gray error bar shows the corresponding constraints for the other object for reference. 
    The yellow curves represent the counterparts whose central BHs are consistent with observed $M_{\rm BH}$ at the target redshift, while the green curves represent the remaining counterparts. 
    Thin curves show the individual counterparts, and thick curves show the median evolution of each subset.
    In the top panels, the gray bands mark the black hole seed mass range adopted in \Abrahma: $3\times 10^{4} \sim 3\times 10^{5}$~\Msun$/h$.
    Because the \gnz\ counterparts are selected from $z=10$ snapshot, some tracks in the middle left panel have $M_{\rm gal}$ below the observational constraint at $z=10.6$.
    The counterparts that match the observed BH masses are characterized by earlier stellar mass assembly and earlier onset of efficient gas accretion. %they enter the efficient accretion phase earlier, with a marked steepening in BH growth around $z\sim12$ (compared with $z\sim10$ for the lower-$M_{\rm BH}$ counterparts), after which $M_{\rm BH}/M_*$ remains approximately constant from $z\sim12$ to 8.
    %A key result is that counterparts matching the observed BH masses do not arise from systematically larger initial seeds; instead, they assemble earlier and enter the efficient accretion phase earlier, with a marked steepening in BH growth around $z\sim12$ (compared with $z\sim10$ for the lower-$M_{\rm BH}$ counterparts), after which $M_{\rm BH}/M_*$ remains approximately constant from $z\sim12$ to 8.
    }\label{fig:galaxy_history}
\end{figure*}

\begin{figure*}[htbp]
    \centering
    \includegraphics[width=1\linewidth]{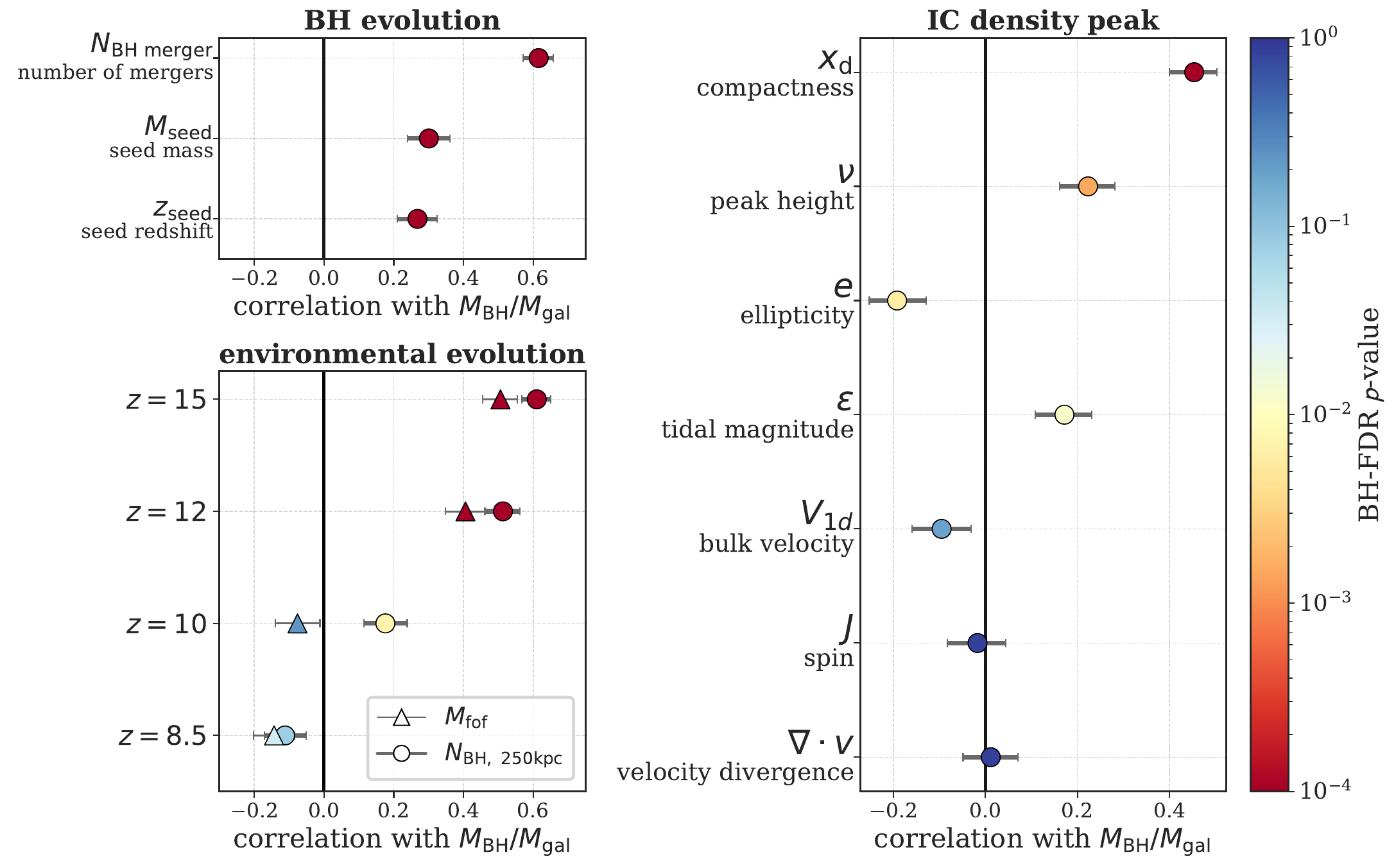}
    \caption{Spearman correlations~(x-axis) between the $z=8.5$ $M_{\rm BH}/M_{\rm gal}$ ratios with a wide range of properties related to BH evolution, environmental evolution, and initial density peak. 
    We use the central BH population hosted by $z=8.5$ central galaxies with $M_*\gtrsim10^{9}$~\Msun. This sample has 247 systems and includes all the \ceers\ counterparts.
    Each marker is colored by BH-FDR adjusted $p$-value, with red indicating more statistically significant correlations. 
    The horizontal error bars indicate 1$\sigma$ confidence intervals for the correlation, estimated via bootstrap resampling. The vertical black line marks zero correlation. 
    \textit{Upper Left:} quantities relating to the BH formation and assembly history i.e. the number of mergers the BH experiences before $z=8.5$~($N_{\rm BH\ merger}$), their seeding mass~($M_{\rm seed}$), and seeding redshift~($z_{\rm seed}$). 
    \textit{Lower Left:} quantities that trace the environmental evolution across time. Specifically, we show the number of neighboring BHs within $250$~kpc (circle) and mass of the host halo $M_{\rm fof}$ (triangle) at  $z=15, 12, 10$, and 8.5. 
    \textit{Right} the correlations with the IC density peak properties.
    The properties are ordered according to the absolute value of their correlation. 
    The compact density peaks tend to produce more massive BHs, as they lead to more efficient BH-BH mergers. 
    }\label{fig:correlation}
\end{figure*}

We use the snapshot at $z=10$ to search for the counterparts of \gnz, which is observed at $z=10.6$, and use the snapshot $z=8.5$ to search for objects similar to \ceers, which was discovered at $z=8.6$. 
In Fig.~\ref{fig:galaxy_scatter}, we present $M_{\rm UV,gal}-M_{\rm gal}$ (upper) and SFR-$M_{\rm gal}$ (lower) for the galaxy population at these two redshifts.
Galaxies within the error bars of the observational constraints of $M_{\rm UV,gal}$, SFR, and $M_{\rm gal}$ are our counterparts, which are colored by the luminosity of their central BH.
%We mark them with dots colored by the luminosity of their central BH (the upper color bar in the rightmost column).  
For \ceers, 
%we convert the $M_{\rm UV, gal}^{\rm intrinsic}$ to $M_{\rm UV,gal}^{\rm obs}$ as the former is not available in the literature. 
the error bars of $M_{\rm UV,gal}^{\rm obs}$ come from the uncertainty in the $M_{\rm UV,gal}^{\rm intrinsic}$--$M_{\rm UV,gal}^{\rm obs}$ scaling relation adopted in \citet{Vogelsberger2020}. 
We have 16 counterparts for \gnz\ and 43 for \ceers. 
As can be seen in Fig.~\ref{fig:galaxy_scatter}, \gnz\ or \ceers-like galaxies are at the high-mass end of the entire galaxy population, while not including the most massive galaxy in either case. The most stringent selection comes from the galaxy magnitude as it has a much smaller error bar compared to SFR and $M_{\rm gal}$. 
For simulated galaxies with \gnz-like stellar masses, almost all of them are within the observed magnitude range and are selected as \gnz\ counterparts. 
But for those with \ceers-like stellar masses, the majority of them are fainter than the observed magnitude range. 
%; only a small fraction among the brighter galaxies in the scatter are selected as \ceers\ counterparts.
%\aklantcomment{Is this what you intended to mean in your last sentence of this para that I commented out? Was not fully clear to me}

%Among galaxies in the same mass range, the \gnz-like galaxies have a typical SFR and $M_{\rm UV, gal}$: the observational space centers around the median $M_{\rm gal, UV}$ and SFR. 
%However, for \ceers, it is brighter than typical galaxies %within its mass range. 

% With these criteria, we identify 16 counterparts for \gnz\ and 43 for \ceers. 
% Full details of the counterpart selection are provided in Appendix~\ref{app:galaxy}. 

%Hence, for \gnz\ and \ceers, observation in general selects the bright galaxies. 

% \aklantcomment{I'm flipping the order of the discussion of the Mbh-Mstar and Mbh-Lbol. The original order was a bit awkward because the reader can already see the BH masses in the Mbh-Lbol panel, but you only discuss them later when talking about Mbh-Mstar. I'd recommend also flipping the upper and lower panels of Figure 5}

\subsubsection{Central BHs of the \gnz\ and \ceers\ counterparts}

Fig.~\ref{fig:Mbh_Mstar} presents the $M_{\rm BH}$--$M_{\rm gal}$ scaling relation for the central BHs hosted by the \gnz\ and \ceers\ galaxy counterparts, along with the full central BH population at the corresponding redshifts. 
%, in the $M_{\rm BH}$--$M_{\rm gal}$ and $L_{\rm bol}$--$M_{\rm BH}$ planes.
%We first focus on the $M_{\rm BH}$--$M_{\rm gal}$ scaling relation~(upper panels) and compare the BH masses of these simulated counterparts against the observed constraints.
The central BHs hosted by our \ceers\ and \gnz\ counterparts (black markers) span more than one order of magnitude in mass. 
For \gnz, 9 of these BHs lie within the observed error bars, corresponding to $\sim56\%$ of the simulated \gnz\ counterparts. These BHs have typical $M_{\rm BH}/M_{\rm gal}$ ratios of $\sim0.001$, consistent with the observed ratio. 
In contrast, for \ceers, the observed $M_{\rm BH}/M_*$ ratio (0.004) is significantly higher than the typical ratio ($\sim0.001$) of the simulated counterparts. 
As a result, only a minority of simulated \ceers\ counterparts ($26\%$) fall within the observationally inferred $M_{\rm BH}$ range, corresponding to 11 BHs. 
These BHs lie along the upper envelope of the $M_{\rm BH}$--$M_{\rm gal}$ plane, and are exactly those which have experienced a larger number of mergers~(see color map on upper panels). 
We will investigate the correlation between $N_{\rm BH merger}$ and $M_{\rm BH}$ further in the next section.  

Next, in Fig.\ref{fig:Lbol_Mbh}, we compare the AGN luminosities of the \gnz\ and \ceers\ counterparts to the observed values on the $L_{\rm bol}$--$M_{\rm BH}$ plane. 
%Next, we compare the AGN luminosities of the \gnz\ and \ceers\ counterparts to the observed values on the $L_{\rm bol}$--$M_{\rm BH}$ plane~(bottom panels of Figure~\ref{fig:Mbh_Mstar}). 
Since the luminosities can vary rapidly on short timescales, we use error bars to indicate their range of values within $\pm25$ Myr around the target snapshot.
%Moreover, considering our capped $\dot{M}_{\rm BH}$ by $2\times$ Eddington while larger Eddington ratio is observed, we mark the counterparts whose peak luminosity $L_{\rm bol, peak}$ within this time window is above $5\times 10^{44}$~erg/s (i.e., 50\% of the observed $L_{\rm bol}$) with circles, and the remaining counterparts with triangles.
As noted in the previous section for the most massive BH, we find that for all \gnz\ and \ceers\ counterparts, the time-averaged AGN luminosities tend to be lower~(by factors $\gtrsim10$) than the observed values. However, if we consider the peak luminosities, 8 out of 43 \ceers\ counterparts (18.6\%) and 2 out of 16 \gnz\ counterparts (12.5\%) produce $L_{\rm bol}$ above 50\% of the observed value, corresponding to $L_{\rm bol}=5\times 10^{44}$~erg/s in both cases.
Recall that part of the reason why we do not fully overlap with the reported \gnz\ luminosity is that our simulations impose an accretion cap of twice the Eddington rate, whereas the observations correspond to an Eddington ratio of $L_{\rm bol}/L_{\rm Edd}\approx5.5$. Additionally, our luminosity fluctuations are likely underestimated by the effective equation of state of the ISM, which artificially smooths density fluctuations in the vicinity of the BH. Overall, we find that comparing the observed luminosities to our simulated counterparts preferentially selects peak phases of the brightest objects within our larger population. Upon considering the observed $L_{\rm bol}$ and $M_{\rm BH}$ together, only 1 \gnz\ (6\%) and 7 \ceers\ (16\%) counterparts match both constraints.

\subsubsection{Evolutionary histories of the \gnz\ and \ceers\ counterparts}

We now examine the evolutionary history of the BHs hosted by the \gnz\ and \ceers\ galaxy counterparts in Fig.~\ref{fig:galaxy_history}. 
For those with $M_{\rm BH}$ consistent with observations (yellow curves), their initial masses span the entire seed mass range ($3\times 10^{4}\sim 3\times 10^{5}\,h^{-1}$~\Msun; see gray band in the top row). This indicates that the initial seed mass is not the dominant factor in assembling BH masses consistent with \gnz\ and \ceers\ at their observed redshifts. Information about the seed mass has been largely erased by subsequent growth in the detectable population. 

Additionally, the growth rate of these massive BHs steepens around $z\sim 12$. This transition corresponds to an enhancement in accretion efficiency. Prior to this redshift, accretion is weak and growth is largely dominated by mergers, until the BH becomes sufficiently massive for the Bondi accretion rate to become efficient. We can also see this transition in the $M_{\rm BH}/M_*$ ratio evolution~(bottom panels of Figure \ref{fig:galaxy_history}). At $z\gtrsim 12$, where mergers dominate BH growth, the $M_{\rm BH}/M_*$ ratio decreases as the galaxy grows faster than the BH. 
Once accretion becomes efficient at $z\lesssim 12$, the BH and the galaxy grow at roughly equal rates, leading to an approximately constant $M_{\rm BH}/M_*$ from $z\sim12-8$. In contrast, for the counterparts that lie below the observed BH masses of \gnz\ and \ceers\ (green curves), this transition occurs at a later redshift of $z\sim 10$. This is because early mergers occur less frequently in these galaxies, so it takes longer for the BH to grow sufficiently massive before Bondi accretion becomes efficient.

Another striking feature of the counterparts that reproduce the observed BH masses is that their progenitor galaxies begin growing earlier than those hosting BHs below the observed masses (middle panels of Figure~\ref{fig:galaxy_history}). 
We will discuss this in more detail in the next section, but this is a direct consequence of 
the high compactness of the initial density peaks in which these galaxies form, which is
one of the key features of the large-scale environment that maximizes BH high-$z$ growth \citep{Ni2022_CR}.
% namely high peak compactness \citep{Ni2022_CR}.
%~(\aklantcomment{cite Ni et al 2022}). 
More specifically, for a set of \gnz\ and \ceers\ galaxies with similar stellar or halo masses, those that begin assembling earlier tend to have higher compactness~(see Figure 2 of \citealt{bhowmick_2022_quasars}). As a result, there is a seemingly counterintuitive consequence of this trend in the $M_*/M_{\rm BH}$ evolution: counterparts with higher BH masses and higher $M_{\rm BH}/M_*$ ratios at later times ($z\lesssim 10$) actually begin with lower $M_{\rm BH}/M_*$ ratios at the earliest times ($z\gtrsim 11$; compare yellow vs green lines in the bottom panels of Figure~\ref{fig:galaxy_history}). 
This is because higher BH masses arise in more compact halos, whose progenitor galaxies already have higher stellar masses at early times, leading to lower $M_{\rm BH}/M_*$ ratios.

\begin{figure*}[!t]
    \centering
    \includegraphics[width=1\linewidth]{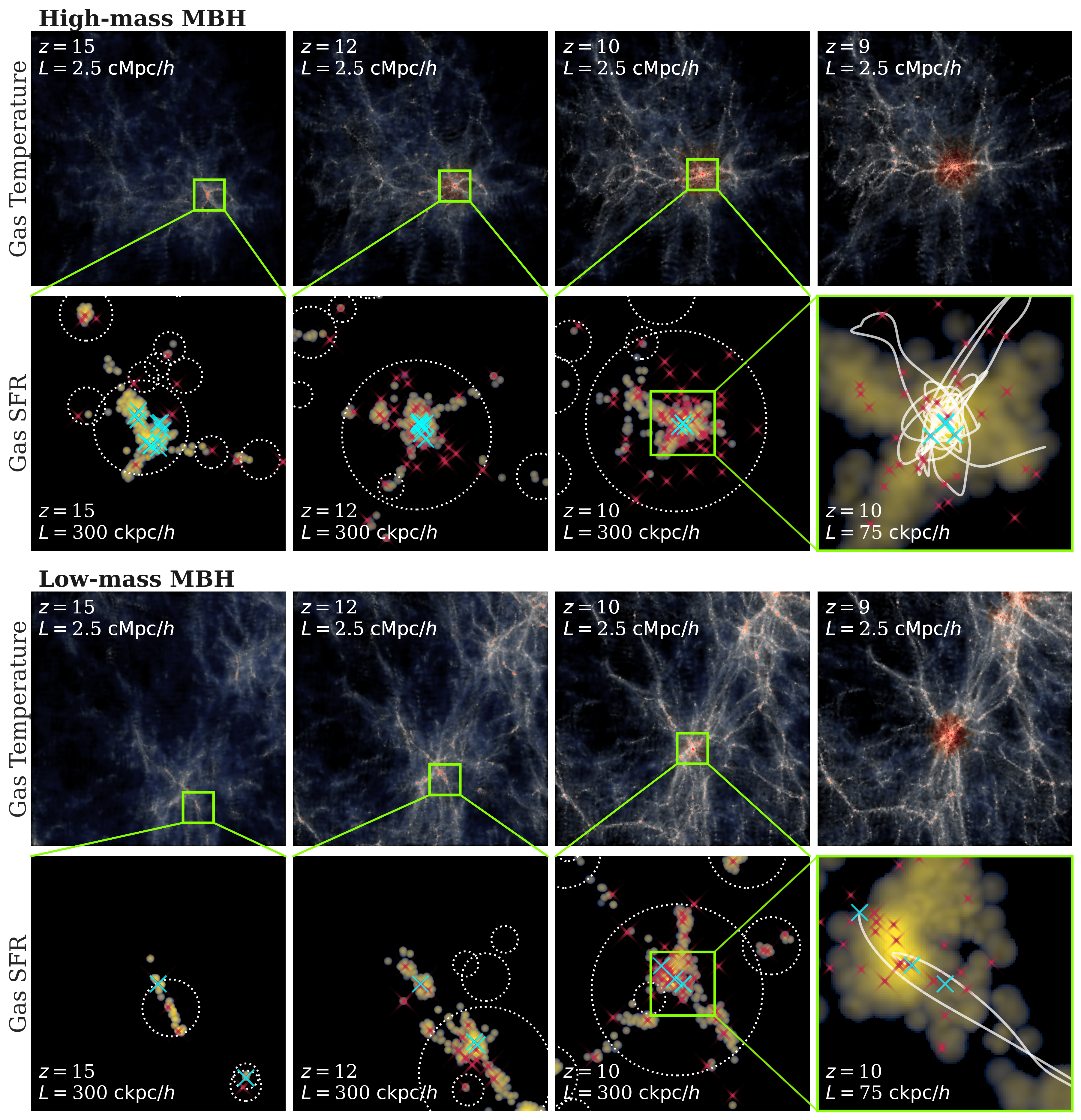}
    \caption{
    Evolution of the host environment of two \ceers\ counterparts from $z=15$ to $z=9$~(left to right). 
    The upper two rows show a system hosting a high-mass BH consistent with the observed JWST measurements ($M_{\rm BH}=8\times 10^{6}$~\Msun\ at $z=8.5$), and the lower two rows show a system hosting a low-mass BH with $M_{\rm BH}$ below the observed measurements ($M_{\rm BH}=4\times 10^{5}$~\Msun\ at $z=8.5$).  The first and third rows show the gas density field in a volume of $2.5\ {\rm cMpc}/h$ per side, colored by gas temperature and centered on the host halo position at $z=9$.
    The green boxes indicate the regions enlarged in the second and fourth rows; these zoom-in panels are centered on the BH progenitors and show the local distribution of star-forming gas (yellow regions) and BHs (red crosses).
    Dashed circles correspond to the virial radii of the halos.
    %We zoom in on regions around the progenitors of the BH counterparts (the green square), and show the star-forming gas (yellow dots) and the BH (red spikes) in the second and fourth panels.       
    %The dashed circles mark the virial radii of the halos. 
    The progenitors of the central BHs of the \ceers\ counterparts are highlighted with cyan crosses. 
    In the rightmost column, we zoom in to a smaller region of $75\ {\rm ckpc}/h$ per side at $z=10$, and plot the BH orbits of the counterpart progenitors. These regions visually illustrate our finding from Figure \ref{fig:correlation}, i.e., the higher BH is produced within a more compact density peak, as this region forms a higher number of seeds with smaller initial separations that can merge within shorter times. 
    }\label{fig:vis_halo_num}
\end{figure*}

\begin{figure*}[!t]
    \centering
    \includegraphics[width=1\linewidth]{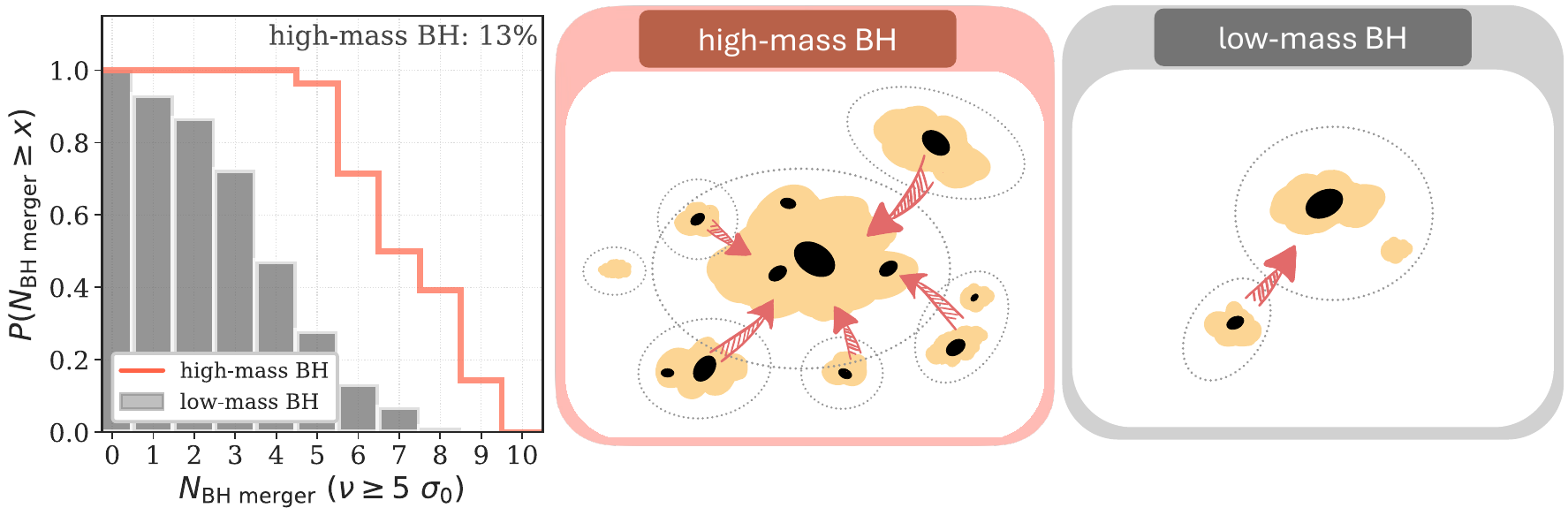}
    \caption{
    \textit{Left:} Cumulative distribution of the number of BH mergers $N_{\rm BH\ merger}$ experienced by $z=8.5$ for systems originating from high-density initial peak (i.e., $\nu \geq 5\ \sigma_{0}$). 
    We divide the sample by the final central BH mass at $z=8.5$: systems with $\log M_{\rm BH}/M_{\odot}\ \geq 6.58$ are classified as high-mass BHs (red), while the rest are classified as low-mass BHs (gray). 
    The threshold $\log M_{\rm BH}/M_{\odot}=6.58$ corresponds to the lower limit of the inferred BH mass for \ceers.  
    There are 13\% of high-density initial peaks that produce high-mass BHs at $z=8.5$.  
    \textit{Middle and Right:} Schematic illustration of typical environments that lead to the formation of a high-mass BH (middle) and a low-mass BH (right). 
    The yellow regions represent star-forming, low-metallicity gas, the black dots mark BH seeds, the dashed circles show halos, and the red arrows indicate subsequent BH merger events. 
    Systems that produce high-mass BHs are embedded in environments with more nearby halos hosting BH seeds, which promotes frequent mergers during the early stages of BH growth. 
    By contrast, low-mass BHs tend to form in more isolated environments, with fewer nearby seeded halos and therefore fewer mergers. 
    %\aklantcomment{I would show a different example compared to the one you already showed in the previous figure. Perhaps even three or four examples of each, since you do have the space for that. Currently, the visualizations feel a bit redundant to the previous figure}. \aklantcomment{Also, upon giving more thought, I'm not actually sure if this figure is actually adding much to the narrative. Upto this point, the paper talks  about compactness being the most important parameter, but here you are showing a correlation with the peak height. I do get the ``12\% fraction argument" , but even so, this figure is redundant with the top left panel of Figure 4. This figure could be more useful if the colorbar showed compactness. Otherwise, we could even remove it completely}
    %The underlying yellow regions represent the star-forming gas, the red crosses represent the BHs, and the white dashed circles represent the virial radius of the halo.
    }
    \label{fig:Nmerger_hist}
\end{figure*}

\begin{figure*}[!t]
    \centering
    \includegraphics[width=0.95\linewidth]{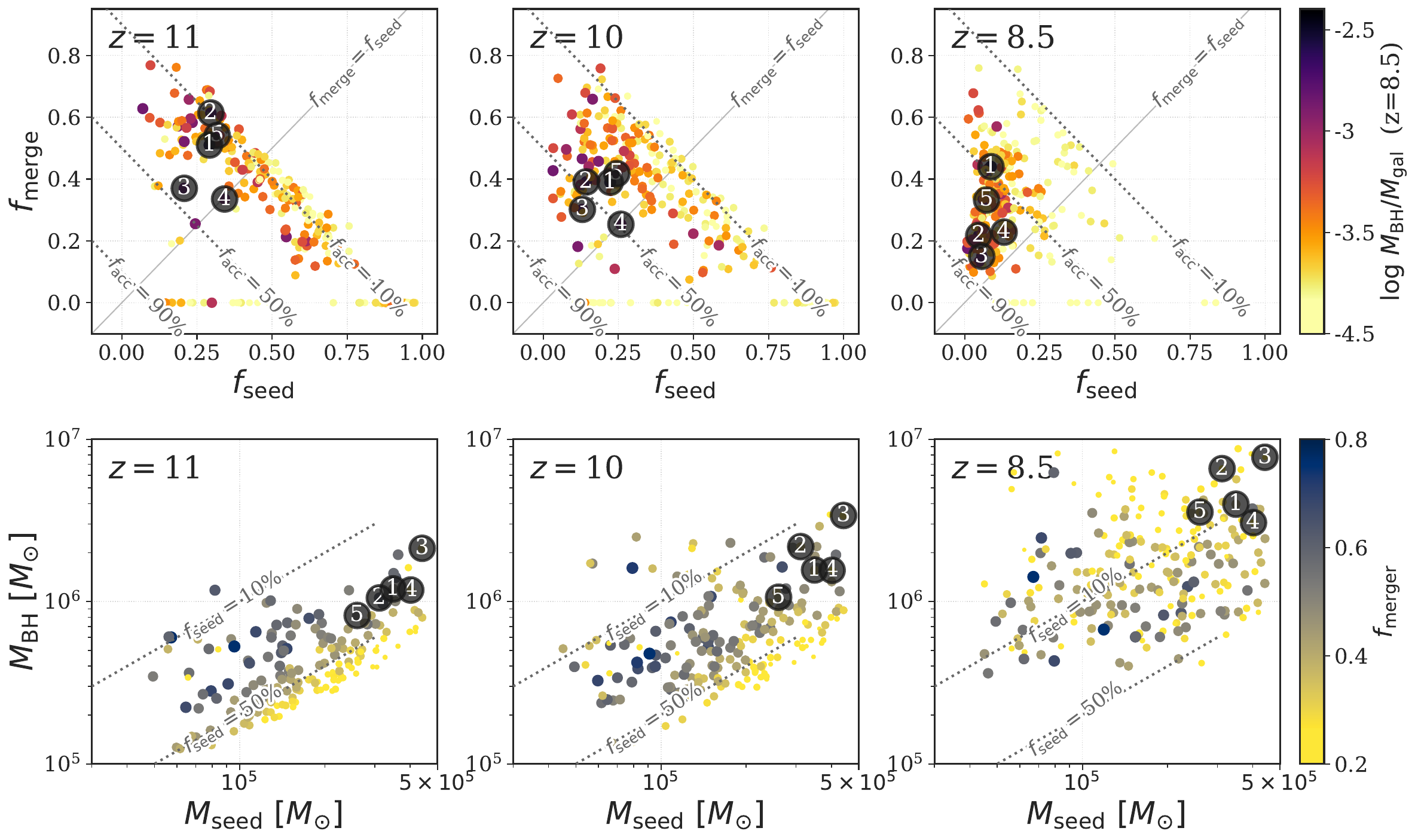}
    \caption{Mass assembly channels of central BHs traced from $z=11$ to the observational epoch $z=8.5$ from left to right.    
    We track the central BHs hosted by the $z=8.5$ central galaxies with $M_{\rm gal}\geq 10^{9}$~\Msun\ back to higher redshift.
    \textit{Top Panels:} Fractional contributions to the BH mass from mergers~($f_{\rm merger}$), plotted against the contribution from the initial seed mass~($f_{\rm seed}$).
    On this $f_{\rm seed}-f_{\rm merger}$ plane, the remaining contribution from gas accretion~($f_{\rm acc}=1-f_{\rm seed}-f_{\rm merger}$) naturally increases as we move from the top-right corner to the bottom-left corner. The three dotted lines correspond to $f_{\rm acc}=10,50$, and $90\%$. 
    We color the markers by the mass ratio $M_{\rm BH}/M_{\rm gal}$ measured at $z=8.5$.
    \textit{Bottom:} $M_{\rm BH}-M_{\rm seed}$ relation at the same redshifts colored by $f_{\rm merger}$. 
    In all the panels, we highlight the evolution of five BHs with the largest $M_{\rm BH}/M_{\rm gal}$ at $z=8.5$ with black circles, and label their rank (`1' denotes the highest $M_{\rm BH}/M_{\rm gal}$).
    For these high-mass BHs, the mass contribution is dominated by mergers at $z=11$. These mergers enhance gas accretion at later times. By $z=8.5$, close to the observed redshift of \ceers, gas accretion becomes the dominant contributor to the BH mass. 
    %This suggests that mergers primarily contribute by enabling more efficient later accretion rather than directly supplying the final mass.
    }\label{fig:mass_contribution}
\end{figure*}

\begin{figure}[htbp]
    \centering
    \includegraphics[width=\linewidth]{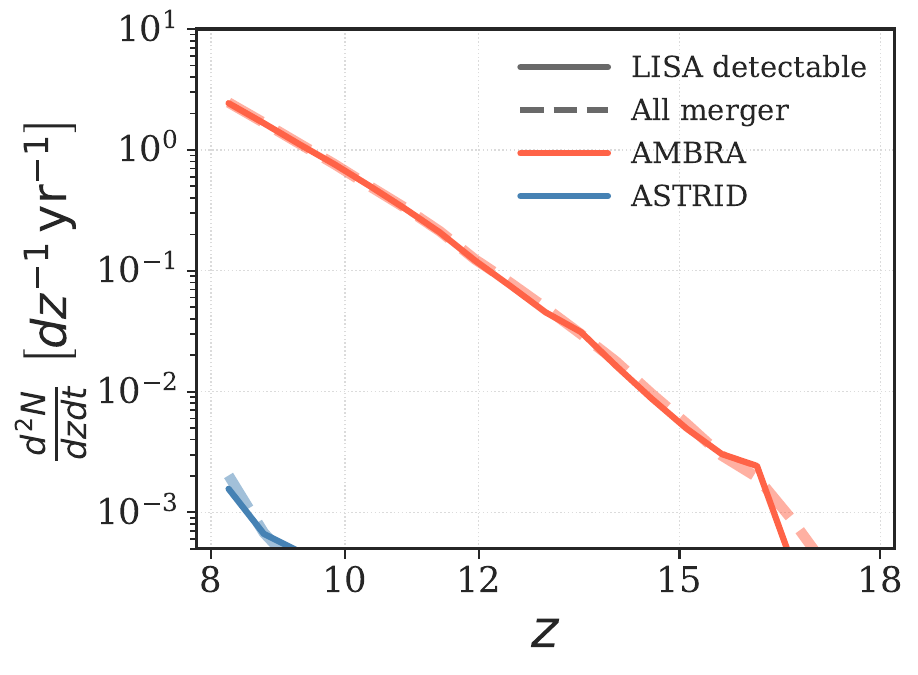}
    \caption{The merger rate of \Abrahma\ (red) and \astrid\ (blue) as a function of redshift. 
    The dashed curves represent all mergers, and the solid curves correspond to those detectable by LISA (SNR$>10$).  
    Compared to \astrid, the overall LISA detection rate in \Abrahma\ is boosted by over three orders of magnitude by $z=8$.
    }
    \label{fig:LISA_snr}
\end{figure}

\subsection{Influence of environment on BH growth}
\label{sec:overmassive}

In the previous section, we showed that even after identifying simulated galaxy counterparts that match multiple galaxy properties ($M_{\rm gal}$, SFR, and $M_{\rm UV}$) of \gnz\ and \ceers, only a subset of them host central BHs massive enough to be consistent with observations. 
This motivates us to investigate which additional properties of the galaxies or their immediate environments are most relevant for maximizing BH growth, particularly in driving high $M_{\rm BH}/M_{\rm gal}$ ratios. 
To address this, we compile a broad set of properties and quantify their influence on $M_{\rm BH}/M_{\rm gal}$ using the Spearman correlation, as presented in Fig.~\ref{fig:correlation}. This statistical analysis uses all central galaxies with $M_{\rm gal}\geq 10^{9}$~\Msun\ at $z=8.5$, the detected redshift for \ceers, yielding a sample of 247 galaxies. Scatter plots for all the properties included in Fig.~\ref{fig:correlation} versus $M_{\rm BH}/M_{\rm gal}$ are shown in Appendix~\ref{app:correlation}.

We first focus on properties that directly trace BH assembly history (upper left panel of Fig.~\ref{fig:correlation}), namely the number of mergers experienced by the central BH prior to $z=8.5$ ($N_{\rm BH\ merger}$), the BH seeding mass ($M_{\rm seed}$), and its formation redshift ($z_{\rm seed}$). 
For $M_{\rm seed}$ and $z_{\rm seed}$, we adopt the seed of the main branch of each BH merger tree.  
As expected, $M_{\rm BH}/M_{\rm gal}$ is positively correlated with all three quantities: BHs that seed earlier, start with larger mass, and undergo more mergers are more likely to become overmassive. Importantly, however, $N_{\rm BH\ merger}$ shows a significantly stronger correlation than either $M_{\rm seed}$ or $z_{\rm seed}$. This indicates that the efficiency of the environment in facilitating BH-BH mergers plays a more dominant role in driving high $M_{\rm BH}/M_{\rm gal}$ ratios than the initial seeding conditions. This trend is also reflected in the pronounced color gradient in the $M_{\rm BH}$-$M_{\rm gal}$ plane shown in Fig.~\ref{fig:Mbh_Mstar}, where color encodes $N_{\rm BH\ merger}$. 
Among the \ceers\ galaxy counterparts, those hosting BH consistent with observed $M_{\rm BH}$ undergo at least five BH mergers before $z=8.5$. 
The counterpart hosting the most massive BH goes through 9 mergers.

The frequency of BH-BH mergers should also be reflected in the surrounding environment. 
In the lower-left panel of Fig.~\ref{fig:correlation}, we examine the correlation between $M_{\rm BH}/M_{\rm gal}$ and two environmental indicators: the number of nearby BHs within 250 kpc ($N_{\rm BH,\ 250,{\rm kpc}}$) and the host halo mass ($M_{\rm fof}$), evaluated at $z=15, 12, 10,$ and $z=8.5$.
Two key trends emerge from this panel. 
First, the correlation with $N_{\rm BH,\ 250,{\rm kpc}}$ is consistently stronger than that with $M_{\rm fof}$, indicating that the presence of nearby BH companions is a more direct predictor of enhanced $M_{\rm BH}/M_{\rm gal}$. These companions increase the likelihood of subsequent mergers, thereby boosting BH growth by $z=8.5$. Second, the correlations are strongest at the highest redshift ($z=15$) and systematically weaken toward lower redshift. This behavior is expected, as BH companions identified at earlier times have a greater probability of merging before $z=8.5$ and contributing to elevated $M_{\rm BH}/M_{\rm gal}$ ratios. Conversely, companions identified at $z=8.5$ are more likely to merge only at later times, and thus show the weakest correlation.

%This indicates that most mergers occur at high redshift ($z\gtrsim 10$), which is consistent with the history plotted in Fig.~\ref{fig:massive_bh_compare}.

While the correlations discussed so far clearly demonstrate the importance of early mergers in maximizing BH mass growth, they do not yet explain why some galaxies,
even at fixed stellar mass, SFR, and UV magnitude,
host more close companions that facilitate these mergers. 
We therefore turn to the initial conditions (ICs) and investigate whether properties of the IC density peaks can predict the emergence of BHs. 
To identify the IC density peak where each $z=8.5$ galaxy forms, we follow \citet{Ni2022_CR} and trace back the dark matter particles in the host halo to the ICs.
We extract a $10$~Mpc region centered on their center of mass, and smooth the overdensity field with a Gaussian kernel of $250$~kpc. We then identify the local density peak within this region and measure several characteristics: peak height ($\nu$), compactness ($x_{\rm d}$), ellipticity ($e$), tidal magnitude (shear scalar; $\epsilon$), bulk velocity ($V_{\rm 1d}$), angular momentum ($J$), and velocity divergence ($\nabla \cdot \mathbf{v}$). Their correlations with $M_{\rm BH}/M_{\rm gal}$ are summarized in the right panel of Fig.~\ref{fig:correlation}. Among these properties, the peak compactness $x_{\rm d}$ emerges as the strongest predictor of $M_{\rm BH}/M_{\rm gal}$, with a Spearman correlation of 0.45, indicating that more compact peaks preferentially give rise to BHs with higher masses. 
Interestingly, a similar dependence on halo compactness has been identified in other BH seeding channels. In the SIDM core-collapse scenario, BH formation naturally occurs in high-concentration halos \citep{Jiang2026}. 
In addition, using a halo-based seeding model similar to fiducial \astrid, \citet{Ni2022_CR} found that high compactness induces rapid BH growth at early epochs.
Taken together, these results suggest that compact proto-halos may be preferential sites for early BH seeding and subsequent rapid growth, even though the underlying seeding physics can differ between models.  

%High-density peaks are often regarded as favorable sites for BH formation \citep[e.g.,][]{Ni2022_CR}, and many small-volume / zoom-in simulations targeting BH formation and growth focus on such regions \citep[e.g.,][]{Bhowmick2025}. However, our results highlight that while high-density peaks can produce massive galaxies, they do not necessarily guarantee the frequent mergers required for rapid BH growth.

%To identify the density peak associated with each host halo, we follow the : we trace back the DM particle belonging to the $z=8.5$ host halo, 

%extract a $10$~Mpc region centered on its center of mass, and smooth the overdensity field with a Gaussian kernel of $250$~kpc. 

%We verified that using a larger smoothing scale (e.g., 1 Mpc) weakens the correlation between the IC density peak and $M_{\rm BH}/M_{\rm gal}$.

In Fig.~\ref{fig:vis_halo_num}, we highlight the connection between the environments and the merger history by showing the evolution of the host environments for two \ceers\ counterparts from $z=15$ to $z=9$. 
The first and second rows correspond to a system hosting a high-mass BH consistent with observational constraints ($M_{\rm BH}=8\times10^{6}~M_{\odot}$ at $z=8.5$). These visualize a highly compact density peak, which seeds multiple BHs in close proximity (red crosses in the second row). 
This configuration promotes rapid early mergers, primarily over $12 \lesssim z \lesssim 15$, with a substantial fraction of these seeds (cyan crosses in the second row) eventually merging into and contributing to the final central BH hosted by the counterpart at $z=8.5$. 
In contrast, the third and fourth rows show a system hosting a counterpart with a lower-mass BH ($M_{\rm BH}=4\times10^{5}~M_{\odot}$ at $z=8.5$, below observational estimates). The initial density peak for this system is significantly less compact. As a result, the seeded BHs are more spatially dispersed (red crosses in the fourth row), reducing the likelihood of mergers. Consequently, only a small fraction of these seeds (cyan crosses in the fourth row) contribute to the growth of the final central BH by $z=8.5$.

Importantly, the correlation with the peak compactness is stronger than that of the peak height $\nu$, which largely determines the halo mass.  
In the left panel of Fig.~\ref{fig:Nmerger_hist}, we show the cumulative distribution of the number of BH mergers $N_{\rm BH\ merger}$ for systems originating from high-density initial peak (i.e., $\nu \geq 5\ \sigma_{0}$). 
Among them, only 13\% produce high-mass BH by $z=8.5$ with $M_{\rm BH}$ above the lower limit of the inferred BH mass for \ceers, which is $\log M_{\rm BH}/M_{\odot}=6.58$. 
And these systems all experience frequent mergers: 97\% of them experience at least five mergers.  
This demonstrates that while high-density peaks can produce massive galaxies, they do not necessarily guarantee the frequent mergers required for rapid BH growth. 
In the middle and right panels of Fig.~\ref{fig:Nmerger_hist}, we present the schematic illustration of typical environments that lead to the formation of high-mass BHs and low-mass BHs. 
Systems that produce high-mass BHs are embedded in environments with more nearby halos hosting BH seeds, which promotes frequent mergers during the early stages of BH growth. 
By contrast, low-mass BHs tend to form in more isolated environments, with fewer nearby seeded halos and therefore fewer mergers.

\subsection{Mass assembly history for high-mass BHs}

Having demonstrated the critical role of early BH-BH mergers in assembling BHs as massive as in \gnz\ and \ceers\ at $z\gtrsim9$, we now track their contribution to the total BH mass assembly across the evolution. %To give a better understanding of how the mergers contribute to the BH mass growth, 
In the top row of Fig.~\ref{fig:mass_contribution}, we consider the $z=8.5$ central BHs hosted by the central galaxies with $M_{\rm gal}\geq 10^{9}$~\Msun, and plot the mass contributions from mergers ($f_{\rm merger}$), the initial seed mass ($f_{\rm seed}$), and gas accretion ($f_{\rm acc}$) at $z=11, 10,$ and $8.5$. 
We also highlight the five systems with the highest $M_{\rm BH}/M_{\rm gal}$ among this sample as black circles.
To estimate the $f_{\rm merger}$, we trace the main branch of each BH merger tree and sum the masses from the secondary BHs in all merger events.
We then divide this total merged mass by the final BH mass at $z=8.5$ to obtain $f_{\rm merger}$.
For $f_{\rm seed}$, we take the seed mass of the BH on the main branch and divide it by the $M_{\rm BH}$ at $z=8.5$.
The remaining contribution from gas accretion is calculated as $f_{\rm acc}=1-f_{\rm seed}-f_{\rm merger}$.
%\aklantcomment{I think we need another sentence here describing how *exactly you calculated these contributions, especially $f_{\rm merger}$ vs $f_{\rm seed}$}. 
%\aklantcomment{Are these 5 systems \ceers\ counterparts}. 
At $z=11$, the majority of the population is still close to its seeding stage, and goes through little gas accretion, with $f_{\rm acc}\sim 10\%$. 
For the more massive BHs (black circles and purple dots), the bulk of their mass is contributed by mergers, with $f_{\rm merger}\sim 50\%$. 
In other words, following seed formation in \Abrahma, the next stage of mass build-up is predominantly driven by BH-BH mergers. This is broadly consistent with the \brahma\ simulations~\citep{Bhowmick2025}. 
From $z=11$ to $z=8.5$, $f_{\rm merger}$ generally declines across this population, while gas accretion becomes the dominant growth channel. 
By $z=8.5$, the most massive BHs are predominantly grown through accretion, with $f_{\rm acc}>50\%$, with a sub-dominant direct merger contribution ($f_{\rm merger}\sim 10$-$40\%$). 
The lower panels of Fig.~\ref{fig:mass_contribution} highlights this trend on the $M_{\rm BH}$-$M_{\rm seed}$ plane colored by $f_{\rm merger}$. 
At $z=11$, the most massive BHs are preferentially merger-dominated. In contrast, the most massive BHs at $z=8.5$ exhibit relatively low $f_{\rm merger}$, indicating a transition to an accretion-dominated growth phase. 
%\aklantcomment{I'm not entirely sure if this last part and the lower panels give us any new insight beyond what is already derived from the upper panels, but it is still useful to visualize these trends on the Mbh-Mseed plane.}
%This suggests that for these $z=8.5$ high-mass MBHs, merger-driven growth occurred at very high redshift, consistent with the correlation between $N_{\rm BH}$ and $M_{\rm BH}/M_{\rm gal}$ shown in the left lower panel of Fig.~\ref{fig:correlation}. 

Importantly, the reduced $f_{\rm merger}$ at $z = 8.5$ does not imply that mergers are unimportant for the assembly of high-mass BHs such as those in \ceers\ within \Abrahma. 
Rather, mergers act as an early catalyst for subsequent accretion-driven growth. By boosting $M_{\rm BH}$ at $z\gtrsim 11$, they enhance later accretion, since the Bondi accretion rate scales as $M_{\rm BH}^{2}$. As a result, a significant fraction of the mass observed at $z=8.5$ is accumulated through accretion that is enabled by the early mergers.

Finally, our prediction for strong merger contribution to the earliest stages of BH growth could be directly tested by future GW observations with the Laser Interferometer Space Antenna (LISA).
Therefore, we show the LISA detection rate for $z\geq 8$ mergers in \Abrahma\ and compare it to \astrid. 
We estimate the LISA signal-to-noise ratio (SNR) following \citet{Wang2025} with an assumption of circular orbits, and classify events as detectable if ${\rm SNR}>10$. 
We generate $10^{4}$ realizations and randomly place a 4-year observational window within the last 100 years prior to coalescence. 
The global detection rate is shown in Fig.~\ref{fig:LISA_snr}. Compared to \astrid\ , \Abrahma\ produces a much higher detection rate: by $z=8$, \astrid\ yields nearly no detectable mergers ($\approx 10^{-3}~{\rm yr}^{-1}$), while \Abrahma\ predicts an event rate of $3.85~{\rm yr}^{-1}$. In other words, a direct consequence of modifying our seed model and reproducing the measured BH masses for \gnz\ and \ceers, is an increase 
over three orders of magnitude in the $z\geq 8$ merger rate from \astrid\ to \Abrahma. Our LISA prediction will provide a direct test of the ``merger-accelerated" early BH assembly scenario seen in the \Abrahma\ simulation, while also highlighting the broader importance of LISA for constraining BH seeding models.

%\st{For further reference, \aklant{the \Abrahma\ event rate at $z>8$ is already comparable to entire \astrid\ event rate of 5.6 ${\rm yr}^{-1}$ for all the mergers it produces down to $z\sim0$~\citep{Wang2025}.} 

%\st{, and consists of $\approx 70\%$ global merger rate across all redshifts} \aklantcomment{I didn't quite understand this phrase}
%This 

\section{Conclusion}\label{sec:conclusion}

In this work, we present the first results of the \Abrahma\ simulation evolved to $z=8$. \Abrahma\ combines the statistical power of ASTRID and the physically motivated gas-based BH seeding models from \brahma\ simulations suite.
Motivated by the JWST discoveries of $z\gtrsim 9$ BH population, we chose one of the most lenient seed models from \brahma\ that allows heavy seeds~($4\times10^4\leq M_{\rm seed} \leq 4\times10^5~M_{\odot}$) to form in every halo with sufficient star-forming and metal-poor gas. 
These seeds could represent end-states of several physical scenarios not resolvable in our simulations, including remnants of stellar collisions in ultra-dense NSCs or rapidly grown PopIII remnants. 
With the same large cosmological volume ($250\ h^{-1}{\rm Mpc}$ per side) and initial conditions as \astrid, \Abrahma\ allows us to investigate the high-redshift BH population across diverse environments, and to directly compare the impact of different seeding models on the early growth of BHs.
%We applied this seed model to a large cosmological box with the exact same volume ($250\ h^{-1}{\rm Mpc}$ per side) and initial conditions as \astrid\ .

Compared to \astrid, \Abrahma\ seeds BHs  far more efficiently and at significantly higher redshift. 
The first BH seeds in \Abrahma\ form at $z \sim 27$, and the seed number density reaches $0.01~\mathrm{d}z^{-1}~\mathrm{Mpc}^{-3}$ by $z \sim 15$. In contrast, \astrid\ forms seeds only after $z \sim 17$.
%The resulting BH number densities at $z \sim 8$ are $\sim 50$ times higher than in \astrid, which only begins forming seeds at $z \sim 17$. 
This early and efficient seeding in \Abrahma~leads to a substantial boost in BH growth, initially driven by BH-BH mergers and subsequently dominated by enhanced gas accretion modeled by Bondi-Hoyle formalism.
As a result, \Abrahma\ is able to produce BHs with masses up to $\sim 10^6~M_{\odot}$ and $\sim 10^7~M_{\odot}$ at $z \sim 10$ and $z \sim 8.5$ respectively, consistent with current BH mass measurements for sources discovered in \gnz\ and \ceers\ with JWST. In contrast, the most massive BHs produced by the \astrid\ simulation are approximately an order of magnitude smaller at these early epochs.

%This is an order of magnitude higher across the mass range $10^{5}\sim 10^{7}$~\Msun\ and the luminosity range $L_{\rm bol}\lesssim 10^{45}$~erg/s by $z=8$.

Importantly, while our large simulation volume contains several galaxy counterparts that match the observed stellar masses, UV luminosities, and star formation rates of \gnz\ (16 objects) and \ceers\ (43 objects), not all of them are able to grow their BHs sufficiently to match current JWST measurements of $M_{\rm BH}$. 
Approximately $56\%$ of the simulated \gnz-like galaxies and $26\%$ of the \ceers-like galaxies attain BH masses consistent with observations. 
An even smaller fraction reaches bolometric luminosities comparable to those observed: $\sim 12.5\%$ for \gnz-like galaxy and $\sim 18.6\%$ for \ceers-like galaxies, and only during the peak phases of their AGN variability.

We find that for this subset of galaxies that reproduces the observed BH properties of \ceers~and \gnz, it is the high compactness of the IC density peak that provides the strongest enhancements to early BH growth. 
Halos with higher compactness produce larger numbers of seeds within close proximity, leading to earlier BH mergers. 
At $z \sim 11$, these mergers are the dominant contributor ($\sim 50\%$) to BH mass assembly in the most rapidly growing systems. 
This merger-driven growth subsequently triggers an earlier onset of efficient gas accretion. 
By $z \sim 8.5$, gas accretion becomes the primary contributor for the most massive BHs consistent with the JWST observations. 

Overall, our new \Abrahma\ simulation demonstrates that if the Universe is able to produce heavy seeds of mass $\sim 10^{5}$~\Msun in sufficient abundance, BH mergers can substantially enhance the BH growth at high redshifts, particularly within highly compact galaxies. This would naturally explain the emergence of $\gtrsim 10^6$-$10^7~M_{\odot}$ BHs at $z\sim8.5$-$11$, consistent with those detected in \gnz\ and \ceers. This scenario is testable with future gravitational wave observations from LISA, as \Abrahma\ predicts a high-redshift ($z > 8$) detection rate of $\sim 3.65$ events per year, which is more than three orders of magnitude higher than that in \astrid.

\begin{acknowledgments}
\section*{Acknowledgments}
YZ and TDM acknowledge the support from the NASA FINESST grant 80NSSC25K0318. 
This work was supported by the National Science Foundation under Cooperative Agreement 2421782 and the Simons Foundation grant MPS-AI-00010515 awarded to the NSF-Simons AI Institute for Cosmic Origins — CosmicAI, https://www.cosmicai.org.
TDM acknowledges funding from NASA ATP 80NSSC20K0519, NSF PHY-2020295, NASA ATP NNX17AK56G, and NASA ATP 80NSSC18K101, NASA Theory grant 80NSSC22K072. AKB and PT acknowledge support from NSF-AST 2510738. LH acknowledges support from the Simons Foundation through the Learning the Universe initiative.
SB was supported in part by Grant 63667 from the John Templeton Foundation.  The opinions expressed in this publication are those of the author(s) and do not necessarily reflect the views of the John Templeton Foundation. SB was supported by NSF AST-2509639. 
\Abrahma~was run on the Frontera facility at the Texas Advanced Computing Center.  
\end{acknowledgments}

\bibliography{reference}{}
\bibliographystyle{aasjournal}

\appendix
\counterwithin{figure}{section}

\section{Resolution test of the BH  seeding history} 
\label{app:resolution}

\begin{figure}[htbp]
    \centering
    \includegraphics[width=0.5\linewidth]{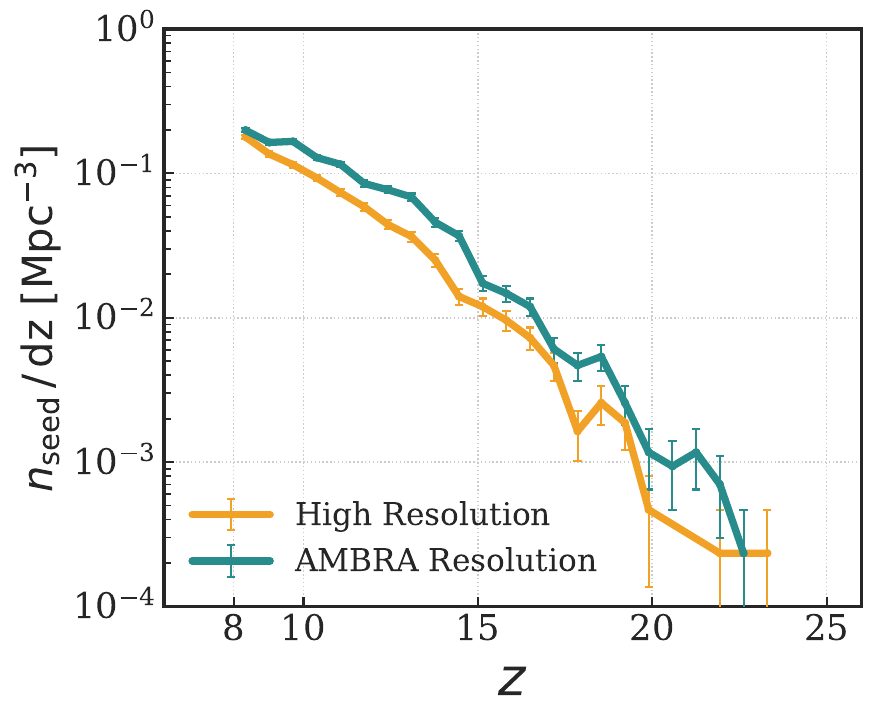}
    \caption{Resolution test for the BH  seeding history in two $12.5\ h^{-1}$~Mpc boxes. 
        We plot the comoving BH  seed number density per unit redshift, $n_{\rm seed}/dz$, as a function of redshift.
        The green curve corresponds to the run with the same resolution as \Abrahma, while the yellow curve shows a run with $\approx 6$ times higher mass resolution.
        Error bars represent the Poisson uncertainties.
        The AMBER-resolution run produces more BH  seeds at high redshift by up to a factor of $\approx 2$, while the two runs converge toward lower redshift. This indicates that the overall redshift dependence of the seeding history is in general robust to resolution.
    }\label{fig:resolution}
\end{figure}

To evaluate the numerical robustness of our BH  seeding model, we perform a resolution convergence test using two simulations in a box of $12.5\ h^{-1}$~Mpc per side. 
One run adopts the same mass resolution as \Abrahma, with gas and dark matter particle masses of $m_{\rm gas}=1.3\times10^{6}\, h^{-1}$~\Msun\ and $m_{\rm DM}=6.7\times10^{6}\,h^{-1}$~\Msun, respectively. 
The gravitational softening length is $\epsilon_{\rm g}=1.5\, h^{-1}{\rm kpc}$.
The second run has a higher resolution, with $m_{\rm gas}=2.0\times10^{5}\, h^{-1}$~\Msun, $m_{\rm DM}=1.1\times10^{6}\, h^{-1}$~\Msun, and $\epsilon_{\rm g}=0.8\, h^{-1}{\rm kpc}$.
This corresponds to a factor of 6 improvement in mass resolution, or a factor of $\sim 1.9$ improvement in spatial resolution.
This higher-resolution setup is chosen to match the heavy-seed model in the BRAHMA simulation suite \citep{Bhowmick2025}.
The two runs are initialized with the same cosmology and the same random seed, so they represent the same realization of the density field, differing only in numerical resolution. 
In both runs, we adopt the same gas-based BH  seeding prescription as \Abrahma: $M_{\rm sfmp}=1.3\times10^{6}\, h^{-1}$~\Msun\ and $M_{\rm h}=2\times10^{8}\, h^{-1}$~\Msun.
The BH seed masses drawn from a power-law distribution over the range $\sim 3\times10^{4}$--$3\times10^{5}\, h^{-1}$~\Msun\ with an index of $-2$. We adopt the same $M_{\rm dyn}=10^{7}h^{-1}$~\Msun\ as in \Abrahma\ for both simulations to prevent numerical heating. 
%A similar calibration was adopted by \citet{chen2024}, where the author used a larger $M_{\rm dyn}/m_{\rm DM}$ ratio than in fiducial \astrid\ when resimulating merging systems from \astrid. 
Keeping $M_{\rm dyn}$ fixed makes the merger rate more robust to resolution, while having little impact on the BH seed number density $n_{\rm seed}$. 

Figure~\ref{fig:resolution} compares the comoving BH  seed number density per unit redshift $n_{\rm seed}/dz$ in these two runs. 
The overall redshift evolution is similar in both cases, indicating that the seeding history is qualitatively robust against resolution changes. 
However, the AMBER-resolution run produces systematically more BH  seeds at high redshift, with an excess of up to a factor of $\sim 2$ relative to the higher-resolution run. 
This offset decreases toward lower redshift.

Overall, this test suggests that the redshift dependence of the seeding history is stable, while the absolute normalization retains a modest resolution dependence in these small-box runs. This level of variation does not qualitatively affect our conclusions regarding the epoch and abundance of BH  seed formation.

\section{The correlation between local environment and $M_{\rm BH}/M_{\rm gal}$}

We present the 2D scatter plot of all properties in Fig.~\ref{fig:correlation}. 
This statistical analysis is based on the 247 central galaxies with $M_{\rm gal}\geq10^{9}$~\Msun\ at $z=8.5$, the redshift where \ceers\ is detected.  
We order the panels based on the absolute value of their Spearman correlation, which is labeled on the lower right corner. The red curves correspond to the median value. 
Among these properties, the number of mergers the BH s experience before the detection epoch, $N_{\rm merger}$, the surrounding BH s within 250 kpc at high redshift ($z\geq 12$), and the compactness of the IC density peaks are strongly correlated with $M_{\rm BH}/M_{\rm gal}$. For an explanation of each property, please refer to Section~\ref{sec:overmassive} and Fig.~\ref{fig:correlation}.

\label{app:correlation}
\begin{figure*}
    \centering
    \includegraphics[width=1\linewidth]{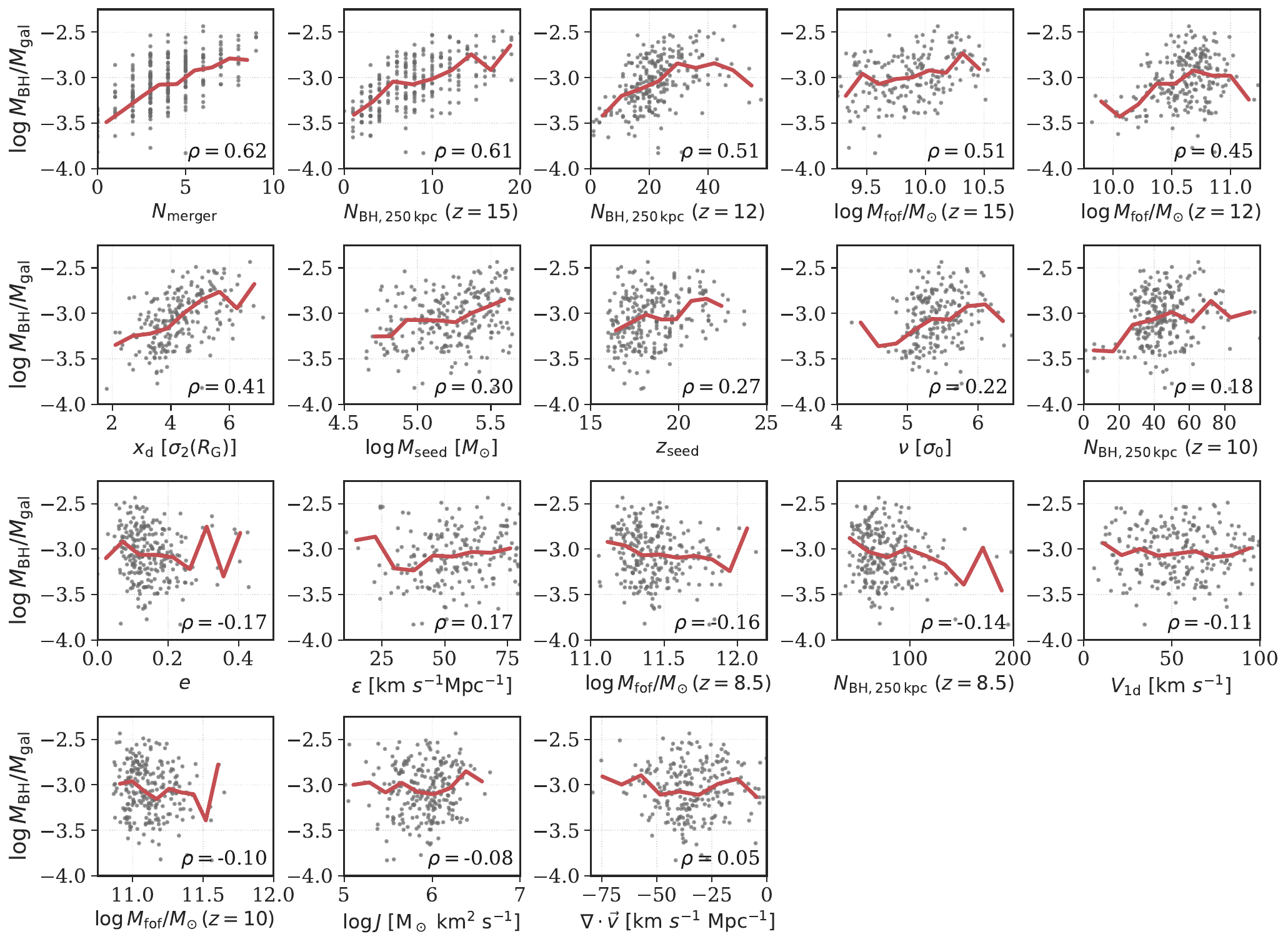}
    \caption{The relation between $M_{\rm BH}/M_{\rm gal}$ and the properties of the local environment. For each panel, we show the results for all central galaxies with mass above $10^{9}$~\Msun\ at $z=8.5$.
    The sample includes 247 galaxies.
    The red curves plot the median value. We label the Spearman correlation in the lower right corner. 
    The panels are ordered based on the absolute values of their correlation. For an explanation of these properties, please refer to Section~\ref{sec:overmassive} and Fig.~\ref{fig:correlation}.
    }\label{fig:corr_scatter}
\end{figure*}

\end{CJK*}
\end{document}